\documentclass{aa}
\usepackage[varg]{txfonts}
\usepackage{makecell}
\usepackage{natbib}
\usepackage[english]{babel} % remove datetime2 complaints about lang not being defined
\usepackage[utf8]{inputenc}
\usepackage[T1]{fontenc}
\usepackage{xspace}
\usepackage[calc]{datetime2}
\usepackage{hyperref}
\usepackage{xcolor}

\definecolor{PineGreen}{HTML}{008B72}
\definecolor{Maroon}{HTML}{AF3235}

\usepackage{savesym}
\savesymbol{tablenum}
\usepackage{siunitx}
\restoresymbol{SIX}{tablenum}
\DeclareSIUnit\day{\text{day}}
\DeclareSIUnit\minute{\text{minute}}

\hypersetup{%
  colorlinks=true,
  linkcolor=blue,
  urlcolor=blue,
  citecolor=blue,
  bookmarksnumbered=true,
  bookmarksopenlevel=1,
  }
\usepackage{etoolbox}
\makeatletter
\patchcmd\@combinedblfloats{\box\@outputbox}{\unvbox\@outputbox}{}{}
\makeatother

\newcolumntype{h}{>{\setbox0=\hbox\bgroup}c<{\egroup}@{}}

\def\FileFont{\tt}

\newcommand\ie{\textit{i.e.}\xspace}
\newcommand\eg{\textit{e.g.}\xspace}
\newcommand\ionelem[2]{#1~{\scshape{#2}}\xspace}
\newcommand\ionline[3]{#1~{\scshape{#2}}~\SI{#3}{\angstrom}\xspace}
\newcommand\sdoaia{\textit{SDO}/AIA\xspace}

\newcommand\kmps{\kilo\meter\per\second}
\renewcommand\exp[1]{\ensuremath{\mathrm{e}^{#1}}}

\RequirePackage[normalem]{ulem}

\DTMnewdatestyle{AASymd}{}
\DTMnewdatestyle{AASmd}{}
\DTMnewdatestyle{AASym}{}
\DTMnewtimestyle{hmsUT}{}
\DTMnewtimestyle{hmUT}{}
\DTMsetdatestyle{AASymd}
\DTMsettimestyle{hmUT}

\newcommand\eispreppro{\texttt{eis\_prep.pro}\xspace}
\newcommand\eisautofitpro{\texttt{eis\_auto\_fit.pro}\xspace}
\newcommand{\eisstudy}[1]{\href{http://solarb.mssl.ucl.ac.uk:8080/SolarB/ShowEisStudy.jsp?study=#1}{#1}}
\newcommand\linefexiimain{\ionline{Fe}{xii}{195.119}}
\newcommand\linefexiisecond{\ionline{Fe}{xii}{186.887}}

\definecolor{plotBlue}{HTML}{004488}
\definecolor{plotRed}{HTML}{BB5566}
\definecolor{plotYellow}{HTML}{DDAA33}
\definecolor{plotGreen}{HTML}{228833}
\definecolor{plotGray}{HTML}{AAAAAA}
\definecolor{plotLightBlue}{HTML}{4F76C1}
\newlength{\plotmarkerheight}
\newlength{\plotlineheight}
\newcommand\plotDetectionMarker{\raisebox{0.05\plotmarkerheight}{\protect\includegraphics[height=\plotmarkerheight]{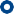}}}
\newcommand\plotApexMarker{\raisebox{0.05\plotmarkerheight}{\protect\includegraphics[height=\plotmarkerheight]{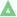}}}
\newcommand\plotLegMarker{\raisebox{0.05\plotmarkerheight}{\protect\includegraphics[height=\plotmarkerheight]{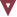}}}
\newcommand\plotReferenceMarker{\raisebox{0.05\plotmarkerheight}{\protect\includegraphics[height=\plotmarkerheight]{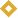}}}
\newcommand\plotAIAMarker{\raisebox{0.05\plotmarkerheight}{\protect\includegraphics[height=\plotmarkerheight]{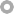}}}
\newcommand\plotAIAEISLine{\raisebox{1.7\plotlineheight}{\protect\includegraphics[height=\plotlineheight]{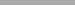}}}
\newcommand\plotAIADetectionLine{\raisebox{1.7\plotlineheight}{\protect\includegraphics[height=\plotlineheight]{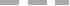}}}

\title{
On the spectroscopic detection of periodic plasma flows in loops undergoing thermal non-equilibrium
}

\titlerunning{Plasma flows in pulsating loops}
\authorrunning{Pelouze et al.}
\date{}

\newcommand{\orcid}[1]{}
\author{%
{Gabriel Pelouze}\inst{\ref{aff:IAS}}\thanks{Present address: Centre for mathematical Plasma Astrophysics, Department of Mathematics, KU Leuven, Celestijnenlaan 200B bus 2400,
3001 Leuven, Belgium.} \orcid{0000-0002-0397-2214}
\and {Frédéric Auchère}\inst{\ref{aff:IAS}} \orcid{0000-0003-0972-7022}
\and {Karine Bocchialini}\inst{\ref{aff:IAS}} \orcid{0000-0001-9426-8558}
\and {Clara Froment}\inst{\ref{aff:Rosseland},\ref{aff:ITA}} \orcid{0000-0001-5315-2890}
\and {Susanna Parenti}\inst{\ref{aff:IAS}}
\and {Elie Soubrié}\inst{\ref{aff:IAS},\ref{aff:IACCC}} \orcid{0000-0001-9295-1863}
}

\institute{%
\label{aff:IAS}{Institut d'Astrophysique Spatiale, CNRS, Univ. Paris-Sud, Université Paris-Saclay, Bât. 121, 91405 Orsay cedex, France}
\\ \email{gabriel.pelouze@kuleuven.be}
\and \label{aff:Rosseland}{Rosseland Centre for Solar Physics, University of Oslo, P.O. Box 1029 Blindern, NO-0315 Oslo, Norway}
\and \label{aff:ITA}{Institute of Theoretical Astrophysics, University of Oslo, P.O Box 1029, Blindern, NO-0315, Oslo, Norway}
\and \label{aff:IACCC}{Institute of Applied Computing \& Community Code, Universitat de les Illes Balears, 07122 Palma de Mallorca, Spain}
}

\abstract%
{
Long-period intensity pulsations were recently detected in the EUV emission of coronal loops, and have been attributed to cycles of plasma evaporation and condensation driven by thermal non-equilibrium (TNE).
Numerical simulations that reproduce this phenomenon also predict the formation of periodic flows of plasma at coronal temperatures along some of the pulsating loops.
}
{
In this paper, we aim at detecting these predicted flows of coronal-temperature plasma in pulsating loops.
}
{
To this end, we use time series of spatially resolved spectra from the EUV imaging spectrometer (EIS) onboard \textit{Hinode}, and track the evolution of the Doppler velocity in loops in which intensity pulsations have previously been detected in images of \sdoaia.
}
{
We measure signatures of flows that are compatible with the simulations, but only in a fraction of the observed events.
We demonstrate that this low detection rate can be explained by line of sight ambiguities, combined with instrumental limitations such as low signal to noise ratio or insufficient cadence.
}
{}


\begin{document}

% for some reason, defining these in the header
% does not work when using the aa class
\renewcommand{\figureautorefname}{Fig.}
\renewcommand{\sectionautorefname}{Sect.}
\renewcommand{\subsectionautorefname}{Sect.}

\maketitle

\section{Introduction} \label{sec:intro}

Understanding the energy transport and heating mechanisms that are able to maintain a million-degree corona around the Sun is a long-standing challenge in astrophysics.
More observational constraints are still needed in order to identify the characteristics of the heating (such as where it is localized and how it changes over time), and to discriminate different heating models.
Long-period intensity pulsations in the extreme-ultraviolet (EUV) emission of coronal loops provide new observables that help constrain the parameters of the heating.
These pulsations were first detected by \citet{AuchereEtAl2014} in images from the \SI{195}{\angstrom} channel of the \textit{Extreme-ultraviolet Imaging Telescope} \citep[EIT:][]{DelaboudiniereEtAl1995} onboard the \textit{Solar and Heliospheric Observatory} \citep[SOHO:][]{DomingoEtAl1995}, and by \citet{FromentEtAl2015} in images from the six coronal channels of the \textit{Atmospheric Imaging Assembly} \citep[AIA:][]{LemenEtAl2012} onboard the \textit{Solar Dynamics Observatory} \citep[SDO:][]{PesnellEtAl2012}.
The pulsations were reported to have periods ranging from \SI{2}{h} to \SI{16}{\hour}, with half of the events occurring in active regions, and \SI{25}{\percent} being visually associated with loops \citep{AuchereEtAl2014, Froment2016}.
% AuchereEtAl2014: 917 events, 54% in AR, 51% of these in loops; Froment2016: 1537 events, 70% in AR, 27% of these in loops.

These pulsations have been interpreted as resulting from Thermal Non-Equilibrium (TNE) \citep{AuchereEtAl2014, FromentEtAl2015, AuchereEtAl2016tne, FromentEtAl2017, FromentEtAl2018}, which can result from of a quasi-constant heating localized near the loops footpoints.
In this case, there may exist no equilibrium between the heating near the footpoints and the radiative losses in the corona (\citealp{AntiochosKlimchuk1991}; \citealp{AntiochosEtAl1999}; \citealp{AntiochosEtAl2000}; \citealp{KarpenEtAl2001}; \citealp{KlimchukLuna2019}; \citealp{Antolin2019}; \citealp{Klimchuk2019}).
As a result, the plasma in the loop undergoes condensation and evaporation cycles (or TNE cycles), during which it periodically changes between a hot and tenuous phase, and a colder and denser phase \citep{KuinMartens1982, MartensKuin1983}.
Enhanced emission in the coronal channels of EIT or AIA occurs during cycles in which the plasma reaches a peak temperature of a few million degrees.
This behavior is well reproduced in one-dimensional hydrodynamic simulations which compute the response of the plasma in a loop to a given heating \citep{KuinMartens1982, MartensKuin1983, KarpenEtAl2001, MullerEtAl2003, MullerEtAl2004, MullerEtAl2005, KarpenEtAl2005, AntolinEtAl2010, XiaEtAl2011, MikicEtAl2013, MokEtAl2016, FromentEtAl2017, FromentEtAl2018}.
In particular, \citet{FromentEtAl2017} were able to convincingly reproduce the intensity and emission measure from one of the events observed with AIA presented in their previous paper \citep{FromentEtAl2015}.

Periodic plasma flows naturally occur in the loop during a cycle, with upflows of hot plasma in both legs during the evaporation phase (simulations of case 1 of \citealp{FromentEtAl2017} predict $\sim\SI{10}{\kmps}$), and strong downflows of cooling plasma that moves towards one of the footpoints during the condensation phase (simulations predict $\gtrsim\SI{50}{\kmps}$ along the loop for plasma at coronal temperatures).
The evaporation phase happens during the minimum of density, which results in very low emission in all the coronal channels of AIA. Therefore we expect that the upflows will be harder to detect.
The downflows start with plasma at coronal temperatures. Depending on the heating parameters, this plasma may then cool down to chromospheric temperatures and form periodic coronal rain showers, or it may be reheated early, thus remaining at coronal temperatures throughout the cycle.

Coronal rain has long been observed in chromospheric and transition region spectral lines, forming blobs-like structures which appear to fall along coronal loops (\citealp{Kawaguchi1970}; \citealp{Leroy1972}; \citealp{Foukal1978}; \citealp{Schrijver2001}; \citealp{DeGroofEtAl2004}; \citealp{deGroofEtAl2005}; \citealp{OSheaEtAl2007}; \citealp{AntolinEtAl2010}; \citealp{AntolinRouppevanderVoort2012}; \citealp{VashalomidzeEtAl2015}).
The formation and dynamics of coronal rain is reproduced with simulations of TNE, both in 1D simulations mentioned above, in 2.5D \citep{FangEtAl2013, FangEtAl2015} and in 3D \citep{MoschouEtAl2015, XiaEtAl2017}.
Coronal rain may also be observed in post-flare loops, where the plasma evaporates and catastrophically cools as a result of the intense transient heating from the flare \citep{ScullionEtAl2016}.
Despite the large number of observations of coronal rain, the periodic nature predicted by simulations of TNE has only been recently observed by \citet{AuchereEtAl2018}.
They report the detection of periodic coronal rain showers observed off-limb in the \SI{304}{\angstrom} channel of AIA.

In this paper, we attempt to detect the flows of plasma at coronal temperatures, which occur regardless of whether coronal rain forms later during the cycle.
While coronal rain is better observed off-limb where it forms distinct blobs \citep{DeGroofEtAl2004, deGroofEtAl2005, AntolinRouppevanderVoort2012}, plasma at coronal temperatures has less distinct structures that could be tracked in plane-of-the-sky images.
Therefore, we attempt to detect these flows on the disk by measuring the Doppler velocity using spectroscopic data from the \textit{EUV Imaging Spectrometer} \citep[EIS:][]{CulhaneEtAl2007} onboard \textit{Hinode} \citep{KosugiEtAl2007}, which can observe lines formed at coronal temperatures.
Depending on the exposure time, EIS can measure velocities with an accuracy ranging from \num{0.5} to \SI{5}{\kmps} \citep{CulhaneEtAl2007, MariskaEtAl2008}.

A number of observational studies have reported average velocities of a few \si{\kmps} in both active regions and the quiet Sun. Transition region lines show systematic redshifts, while coronal lines show blueshifts \citep{SandlinEtAl1977, PeterJudge1999, TeriacaEtAl1999, DadashiEtAl2011}.
While this may change the absolute Doppler velocities in coronal loops, it should not affect the amplitude of the velocity variations.
Furthermore, evidence of flows associated with condensation in loops have been reported both at transition region and coronal temperatures \citep{OSheaEtAl2007, KamioEtAl2011, OrangeEtAl2013}.
However, the periodic nature of the flows at coronal temperatures has never been observed.

To get an estimation of the expected velocity variations, we look at a simulation of TNE cycles without formation of coronal rain performed by \citet{FromentEtAl2017} using the realistic geometry from a loop in which \citet{FromentEtAl2015} detected long-period intensity pulsations.
In this simulation, the velocity along the loop leg changes from \num{-10} to \SI[retain-explicit-plus]{+55}{\kmps} over a cycle, hence an amplitude of \SI{65}{\kmps}.
Simulations of coronal rain formed during TNE cycles predict velocities as high as several \SI{100}{\kmps} \citep[\eg][]{MullerEtAl2004, AntolinEtAl2010, JohnstonEtAl2019}.
However, these fast flows only occur when the plasma reaches transition region or chromospheric temperatures, and should therefore not be observed in coronal-temperature lines.
Simulations from \citet{FromentEtAl2018} show that the velocity profiles are similar for cycles both with and without coronal rain outside of the coronal rain phase, with values consistent with \citet{FromentEtAl2017}.
They further show that the coronal rain phase only occupies a few percent of the cycle period.
Hence, cycles with and without coronal rain should have similar velocity signatures in spectral lines formed at coronal temperatures.

However, accurately predicting the Doppler velocity signature of such flows is challenging, as it is affected by the projection and integration along the line of sight (LOS).
The angle between tho loop and the LOS could be estimated using magnetic field extrapolation and loop tracing methods in EUV images (see \eg \citealp{WarrenEtAl2018} and references therein).
The result would be different for every observed loop as it depends on the loop geometry and the LOS position.
For the loop simulated by \citet{FromentEtAl2017}, the velocity projected along the LOS of AIA or EIS changes from \num{-5} to \SI[retain-explicit-plus]{+30}{\kmps} over a cycle in the loop leg.

Fully understanding the effects of LOS integration would require forward modelling from 2D or 3D simulations of TNE.
How LOS integration affects the time lag signature of TNE cycles has been studied by \citet{WinebargerEtAl2016} using 3D simulations, but no such work exists for the velocities.
We approximate the effects of LOS integration by supposing that the plasma outside of the loop is on average at rest.
Using Monte-Carlo simulations (described in \autoref{subsec:discussion_amplitude}), we show that under this assumption, the Doppler velocity measured with EIS depends on the velocity along loop, the angle between the loop and the LOS, and on the ratio of the intensities emitted by plasma in the loop ($I_\text{loop}$) and elsewhere on the LOS ($I_\text{LOS}$).
Simulations predict that the AIA \SI{193}{\angstrom} intensity ($\sim\SI{1.6}{MK}$) emitted by a single loop strand can vary by a factor ranging from 10 \citep[figure 6]{WinebargerEtAl2016} to 100 \citep[figure 11]{FromentEtAl2017} during a TNE cycle.
The coronal-temperature emission from the loop is therefore negligible at the intensity minimum of the cycle, such that $I_\text{min} \simeq I_\text{LOS}$. Therefore, we can approximate $I_\text{loop} / I_\text{LOS} \simeq (I_\text{max} - I_\text{min}) / I_\text{min}$.
Taking into account the projection and integration along the line of sight, the measured Doppler velocity variations could range from about \num{3} to \SI{30}{\kmps}.
These velocities are comparable to the typical accuracy of EIS, and should therefore be detectable.

We also take advantage of the EIS spectroscopic data to track the evolution of the density in some of the pulsating loops, and compare it to the simulations.

In addition to the periodic flows of plasma at coronal temperatures, multidimensional simulations by \citet{FangEtAl2013, FangEtAl2015} and \citet{XiaEtAl2017} predict that coronal rain should be accompanied by simultaneous counter-streaming flows occurring in adjacent field lines.
If such flows occur at coronal temperatures, they should result in a periodic broadening of the spectral lines during the TNE cycles.
However, the line width also depends on the temperature and the presence of downflows in the loop, which both change over a cycle.
Separating these different contributions would require an extensive analysis which is outside of the scope of this paper.

In \autoref{sec:finding_data}, we describe the search for sets of EIS data suitable for this analysis.
In \autoref{sec:method}, we present the method used to analyze these datasets and to measure the velocity and density.
In section \autoref{sec:results}, we present the results from four datasets, two of which have velocities compatible with the simulations, despite being at the detection limit.
We discuss these results in \autoref{sec:discussion}, and summarize them in \autoref{sec:conclusion}.

\section{Finding appropriate datasets} \label{sec:finding_data}

% tab:datasets {{{
\renewcommand{\theadalign}{tc}
\renewcommand{\cellalign}{tr}
\begin{table*}
\footnotesize
\caption{
  Main characteristics of the 11 EIS datasets that correspond to the observation of loops undergoing long-period intensity pulsations.
  For datasets composed of rasters with different EIS study types, the raster count, exposure time, and slit width are given for each study on different lines.
  \label{tab:datasets}}
\begin{tabular}{chhcchccchccccc}
\hline\hline
  {} &
  {ID} &
  {Event} &
  {Start date} &
  \shortstack{Duration \\ ~[h]} &
  {Amplitudes} &
  \shortstack{Intensity \\ contrast\tablefootmark{(a)}} &
  \shortstack{Pulsation \\ period [h]} &
  \shortstack{Nr. of \\ periods} &
  \shortstack{Nr. of \\ rasters} &
  \shortstack{Rasters per \\ period\tablefootmark{(b)}} &
  \shortstack{EIS \\ study ID} &
  \shortstack{Raster \\ count} &
  \shortstack{Exposure \\ time [s]} &
  \shortstack{Slit width \\ ~[\arcsec]} \\
\hline
\makecell{1}	& \makecell{2}	& \makecell{211\_20110902}	& \makecell{2011-09-03 10:56:15}	& \makecell{40.1}	& \makecell{50\% (50\%)}	& \makecell{50\%}	& \makecell{5.8}	& \makecell{6.9}	& \makecell{240}	& \makecell{34.8}	& \makecell{\eisstudy{461}}	& \makecell{240}	& \makecell{9}	& \makecell{2}	\\
\makecell{2}	& \makecell{3}	& \makecell{193\_20120603}	& \makecell{2012-06-08 03:03:27}	& \makecell{28.7}	& \makecell{35\% (35\%)}	& \makecell{35\%}	& \makecell{4.9}	& \makecell{5.9}	& \makecell{178}	& \makecell{30.2}	& \makecell{\eisstudy{485} \\\eisstudy{486}}	& \makecell{166 \\12}	& \makecell{3 \\ 3}	& \makecell{2 \\2}	\\
\makecell{3}	& \makecell{7}	& \makecell{171\_20120730}	& \makecell{2012-07-30 05:57:42}	& \makecell{123.9}	& \makecell{75\% (35\%)}	& \makecell{35\%}	& \makecell{10.0}	& \makecell{12.4}	& \makecell{793}	& \makecell{64.0}	& \makecell{\eisstudy{485}}	& \makecell{793}	& \makecell{3}	& \makecell{2}	\\
\makecell{4}	& \makecell{10}	& \makecell{171\_20130515}	& \makecell{2013-05-15 21:30:55}	& \makecell{123.7}	& \makecell{60\% (12\%)}	& \makecell{12\%}	& \makecell{8.2}	& \makecell{15.1}	& \makecell{378}	& \makecell{25.0}	& \makecell{\eisstudy{461} \\\eisstudy{480} \\\eisstudy{428}}	& \makecell{376 \\1 \\1}	& \makecell{9 \\ 15 \\ 45}	& \makecell{2 \\1 \\1}	\\
\makecell{5}	& \makecell{8}	& \makecell{171\_20130821}	& \makecell{2013-08-22 18:59:34}	& \makecell{34.6}	& \makecell{25\% (15\%)}	& \makecell{15\%}	& \makecell{4.9}	& \makecell{7.1}	& \makecell{194}	& \makecell{27.3}	& \makecell{\eisstudy{485}}	& \makecell{194}	& \makecell{3}	& \makecell{2}	\\
\makecell{6}	& \makecell{5}	& \makecell{171\_20140618}	& \makecell{2014-06-21 15:26:13}	& \makecell{81.1}	& \makecell{30\% (20\%)}	& \makecell{20\%}	& \makecell{4.7}	& \makecell{17.3}	& \makecell{131}	& \makecell{7.6}	& \makecell{\eisstudy{461}}	& \makecell{131}	& \makecell{9}	& \makecell{2}	\\
\makecell{7}	& \makecell{6}	& \makecell{171\_20140724}	& \makecell{2014-07-27 02:29:14}	& \makecell{74.6}	& \makecell{15\% (10\%)}	& \makecell{10\%}	& \makecell{6.3}	& \makecell{11.8}	& \makecell{184}	& \makecell{15.6}	& \makecell{\eisstudy{461}}	& \makecell{184}	& \makecell{9}	& \makecell{2}	\\
\hline
\makecell{8}	& \makecell{1}	& \makecell{171\_20101102}	& \makecell{2010-11-03 21:15:34}	& \makecell{5.8}	& \makecell{30\% (20\%)}	& \makecell{20\%}	& \makecell{3.9}	& \makecell{1.5}	& \makecell{60}	& \makecell{40.0}	& \makecell{\eisstudy{358}}	& \makecell{60}	& \makecell{20}	& \makecell{2}	\\
\makecell{9}	& \makecell{4}	& \makecell{171\_20121102}	& \makecell{2012-11-04 20:42:29}	& \makecell{13.5}	& \makecell{30\% (5\%)}	& \makecell{5\%}	& \makecell{5.8}	& \makecell{2.3}	& \makecell{88}	& \makecell{38.3}	& \makecell{\eisstudy{358}}	& \makecell{88}	& \makecell{20}	& \makecell{2}	\\
\hline
\makecell{10}	& \makecell{9}	& \makecell{193\_20120107}	& \makecell{2012-01-10 15:35:49}	& \makecell{22.0}	& \makecell{15\% (15\%)}	& \makecell{15\%}	& \makecell{6.3}	& \makecell{3.5}	& \makecell{6}	& \makecell{1.7}	& \makecell{\eisstudy{437}}	& \makecell{6}	& \makecell{60}	& \makecell{1}	\\
\makecell{11}	& \makecell{11}	& \makecell{193\_20140809}	& \makecell{2014-08-09 15:37:25}	& \makecell{116.8}	& \makecell{30\% (30\%)}	& \makecell{30\%}	& \makecell{9.4}	& \makecell{12.4}	& \makecell{21}	& \makecell{1.7}	& \makecell{\eisstudy{428} \\\eisstudy{480}}	& \makecell{12 \\9}	& \makecell{45 \\ 15}	& \makecell{1 \\1}	\\

\hline
\end{tabular}
\tablefoottext{a}{Contrast of the intensity pulsations in the AIA~\SI{193}{\angstrom} channel during the EIS observations, estimated visually.} \\
\tablefoottext{b}{Average number of rasters per period, \ie number of rasters divided by the number of periods.}
\end{table*}
% }}}

In order to detect the predicted pulsations in velocity, it is required to observe the same active region continuously during several pulsation periods, with several measurements per period.
For periods around \SI{10}{\hour}, this translates into several days of observation.
We use data from \textit{Hinode}/EIS, which can acquire spatially resolved spectra (rasters) by scanning a slit across the field-of-view (FOV).

We consider 3181 long-period intensity pulsation events that were detected with AIA between 2010 and 2016 by \citet{Froment2016}, using the method presented in \citet{FromentEtAl2015}.
For each of these events, we systematically search the EIS database (\url{http://sdc.uio.no}) for sets of rasters such that:
\begin{enumerate}
\item the FOV of each raster intersects with the region where pulsations are detected with AIA data;
\item the FOV is wider than \SI{55}{\arcsec} to exclude narrow rasters and sit-and-stare studies;
\item the dataset duration is longer than three pulsation periods;
\item the gaps between the rasters are not too long nor too frequent (this last criterion is estimated qualitatively).
\end{enumerate}
We do not constrain other raster parameters such as the exposure time, the slit width, or the step between consecutive slit positions.

Overall, 11 datasets are found, and their characteristics are presented in \autoref{tab:datasets}.
In addition to the parameters of the EIS observations, this table shows the period of the intensity pulsations that were detected with AIA, and an estimation of their amplitude during the EIS observing period.
We quantify the amplitude of the pulsations with the contrast of the maximum to the minimum intensity ($ (I_\text{max} - I_\text{min}) / I_\text{min}$), measured in the \SI{193}{\angstrom} band of AIA over a cycle.
As argued in the introduction, this provides a reasonable approximation of the contrast between the loop and the background emission.

These datasets can be divided into three categories.
The first and largest category (datasets 1 to 7) contains datasets with a good cadence (ten or more rasters per pulsation period), but short exposure times (less than \SI{10}{\second}) and therefore low signal-to-noise-ratio (SNR). % (The cadence is high enough to detect pulsations if they exist, but the SNR is too low to get an accurate velocity measurement.)
The second category (datasets 8 and 9) also contains datasets with a good cadence, composed of rasters that have longer exposure times (thus better SNR), but narrow FOVs (\SI{60}{\arcsec} along the X axis, \ie perpendicular to the slit) and short total observing time (1.5 and 2.3 pulsation periods respectively).
Finally, the last category (datasets 10 and 11) contains rasters with the highest SNR, but a very low cadence (about one raster per pulsation period). % (Although the SNR allows for accurate velocity measurements, the cadence is not high enough to detect possible pulsations.)
While none of these datasets fulfill all the criteria required to detect with certainty the expected pulsations, those with both a good SNR and large-amplitude intensity pulsations should allow for the detection of the predicted velocities.

\section{Analyzing time series of EIS rasters} \label{sec:method}

We measure the intensity and Doppler velocity using the \linefexiimain line.
This line is formed at a temperature of \SI{1.6}{MK}, which is attained during the cooling phase of most simulated cycles \citep{FromentEtAl2017, FromentEtAl2018}.
It is also one of the brightest lines observed by EIS \citep{YoungEtAl2007}, which will help maximize the SNR, as well as the main contributor to the AIA \SI{193}{\angstrom} band \citep{BoernerEtAl2012}, which will allow for easy comparison with AIA observations.
When the \linefexiisecond line is available, we derive the density from the \ionelem{Fe}{xii} \num{186.887}~/~\SI{195.119}{\angstrom} ratio, which is sensitive to electron number density in the \num{e14}–\SI{e18}{m^{-3}} range \citep{YoungEtAl2007, YoungEtAl2009}, and covers the expected loop densities of \num{e14}–\SI{e15}{m^{-3}} \citep{FromentEtAl2017, FromentEtAl2018} for cycles where the plasma remains at coronal temperatures.

\paragraph{Data preparation and line fitting}

Each EIS raster is first prepared into level~1 data using the \eispreppro routine from SolarSoft \citep{FreelandHandy2012}.
We then fit gaussians to the \linefexiimain and \SI{186.887}{\angstrom} lines using the SolarSoft routine \eisautofitpro, which allows us to derive intensity and velocity maps for these two lines.
The \SI{195.119}{\angstrom} line is blended with a weaker \ionelem{Fe}{xii} line at \SI{195.179}{\angstrom}. We fit this feature using two gaussians that share the same width and have a fixed wavelength separation of \SI{0.06}{\angstrom} \citep{YoungEtAl2009}.
The wing of the \SI{186.887}{\angstrom} line contains a weaker line at \SI{186.976}{\angstrom}, which \citet{BrownEtAl2008} suggested could be a \ionelem{Ni}{xi} transition. We use two independent gaussians to fit these lines.
\linefexiisecond is also blended with a weak \ionelem{S}{xi} line at \SI{186.839}{\angstrom}. Although the contribution from this line is difficult to quantify, \citet{YoungEtAl2009} report that it is below 10\%, and only has a small effect on the resulting densities. We therefore decide not to correct for its contribution.
Each of the aforementioned electronic transitions results in two distinct contributions: a Doppler-shifted component from plasma flowing in the loop, and non-shifted component from plasma at rest elsewhere on the line of sight.
We demonstrate in \autoref{subsec:discussion_amplitude} that the velocity in the loop is best retrieved when fitting a single gaussian to each transition.

In addition, we verify whether the fit results are significantly modified when correcting for the effect described by \citet{KlimchukEtAl2016}, which is that the spectral intensity integrated within a wavelength bin is different than the intensity at the center of this bin.
We tried to correct the spectral intensities using the Intensity Conserving Spectral Fitting method \citep{KlimchukEtAl2016}.
This marginally changed the fit results (typically \SI{0.05}{\kmps} for the line position, and \SI{0.1}{\percent} for the integrated intensity), hence we did not correct the data for this effect.

\paragraph{Spatial coalignment}

In order to get accurate pointing information, we coalign all EIS rasters with AIA \SI{193}{\angstrom} images using the method presented in \citet{PelouzeEtAl2019eis_roll}. This allows to correct for the pointing offset, the instrument roll, and the spacecraft jitter.
All maps are then converted into Carrington heliographic coordinates in order to compensate for the effect of solar rotation.
The differential rotation is corrected using the rotation rate $\Omega(\phi)$ measured by \citet{Hortin2003} for the \SI{195}{\angstrom} channel of EIT:
\begin{equation}
\Omega(\phi) = 14.50 - 2.14 \sin^2\phi + 0.66 \sin^4\phi \ [\si{\degree\per\day}]
\end{equation}
(where $\phi$ is the heliographic latitude).

\paragraph{Velocity measurement}

The velocity is derived from the centroid of the fitted \linefexiimain line. We adopt the convention that positive velocities correspond to spectral redshifts, \ie plasma moving away from the observer.

\paragraph{Correction of orbital effects}

The velocities measured with EIS are affected by thermoelastic deformations of the instrument caused by the orbit of \textit{Hinode}: over the 98-minutes orbit, the position of the spectrum on the detector drifts periodically, which introduces time-dependant velocity variations of up to \SI{70}{\kilo\meter\per\second} \citep{BrownEtAl2007, KamioEtAl2010, EISSWN5}.
The measured absolute velocities can therefore change significantly between different rasters, or even within a raster if the raster duration is comparable to the orbital period.
Two different methods can be used to correct for this spectrum drift. In the first method, the quiet Sun is used as a reference to estimate the drift directly from the data \citep[see \eg][]{BrownEtAl2007, MariskaEtAl2007, YoungEtAl2012}.
The second method was developed by \citet{KamioEtAl2010}, and uses an empirical model to predict the spectral drift from EIS housekeeping data.
The \citet{KamioEtAl2010} correction is applied within the SolarSoft routine \eisautofitpro, but it does not fully correct for the spectral drift, leaving residuals of about \SI{5}{\kilo\meter\per\second}.

A second correction is thus needed in order to detect the pulsations of a few kilometers per second.
The high cadence datasets (1~to~7) are composed of rasters with narrow FOVs ($\ang{;;162}\times\ang{;;152}$ to $\ang{;;180}\times\ang{;;152}$) centered on the active region, which contain little to no quiet Sun.
% numbers: 152"×162" for studies 485 and 486, 152"×180" for study 461. Studies 428 and 480 have wider FOVs but only 1 raster from each so nobody cares.
For these rasters, we use as a reference the \linefexiimain velocity averaged in the region over which the FOVs of \SI{95}{\percent} of the rasters overlap, and set it to \SI{0}{\kmps}. Because the duration of these rasters is short compared to the orbit (\num{2.7} to \SI{4.5}{minutes} \textit{vs.} \SI{98}{minutes}), it is acceptable to use a common velocity reference for all slit positions.
The other datasets (8 to 11) have tall FOVs (\ang{;;368} to \ang{;;512}), which usually contain quiet Sun at the North or South of the active region. In this case, we compute the average velocity in the quiet Sun region for each slit position, and use it as a reference. We correct the intrinsic quiet Sun velocity using the method described by \citet{YoungEtAl2012}, and the average shift of \SI{-2.4}{\kmps} reported by \citet{DadashiEtAl2011} for \ionelem{Fe}{xii}.

\paragraph{Density measurement}
The density is measured through the \ionelem{Fe}{xii} \num{186.887} / \SI{195.119}{\angstrom} lines ratio, which is sensitive to density variations in the \num{e8}–\SI{e12}{cm^{-3}} range \citep{YoungEtAl2007, YoungEtAl2009}.
We derive the densities from this line ratio using Chianti version 8 \citep{DereEtAl1997-chianti1, DelZannaEtAl2015-chianti14v8}, and assuming the temperature of \SI{1.6}{\mega\kelvin}, the peak formation temperature of \ionelem{Fe}{xii}.
The density could not be measured in datasets~1, 4, 6, and 7, because they do not contain the \linefexiisecond line.

\paragraph*{}\hspace{\parindent} % section conclusion
After applying the previous steps, we obtain series of corrected intensity, velocity, and (when possible) density maps for each of the datasets listed in \autoref{tab:datasets}.

\section{Results} \label{sec:results}

We present the analysis of four of the datasets presented at the end of \autoref{sec:finding_data}:
datasets~1 and 8 for which the SNR and the amplitude of the intensity pulsations are large enough to allow for the detection of velocity variations that are compatible with the expected pulsations (\autoref{subsec:pulsations});
datasets~2 and 11 in which no velocity pulsations could be detected either due to a low SNR, or to insufficient cadence (\autoref{subsec:no_pulsations}).
No pulsations in velocity are detected in the other datasets.

\subsection{Datasets with velocities consistent with the expected pulsations}\label{subsec:pulsations}

% dataset 1 {{{

\begin{figure} % fig:ds1_maps
\includegraphics[width=\columnwidth]{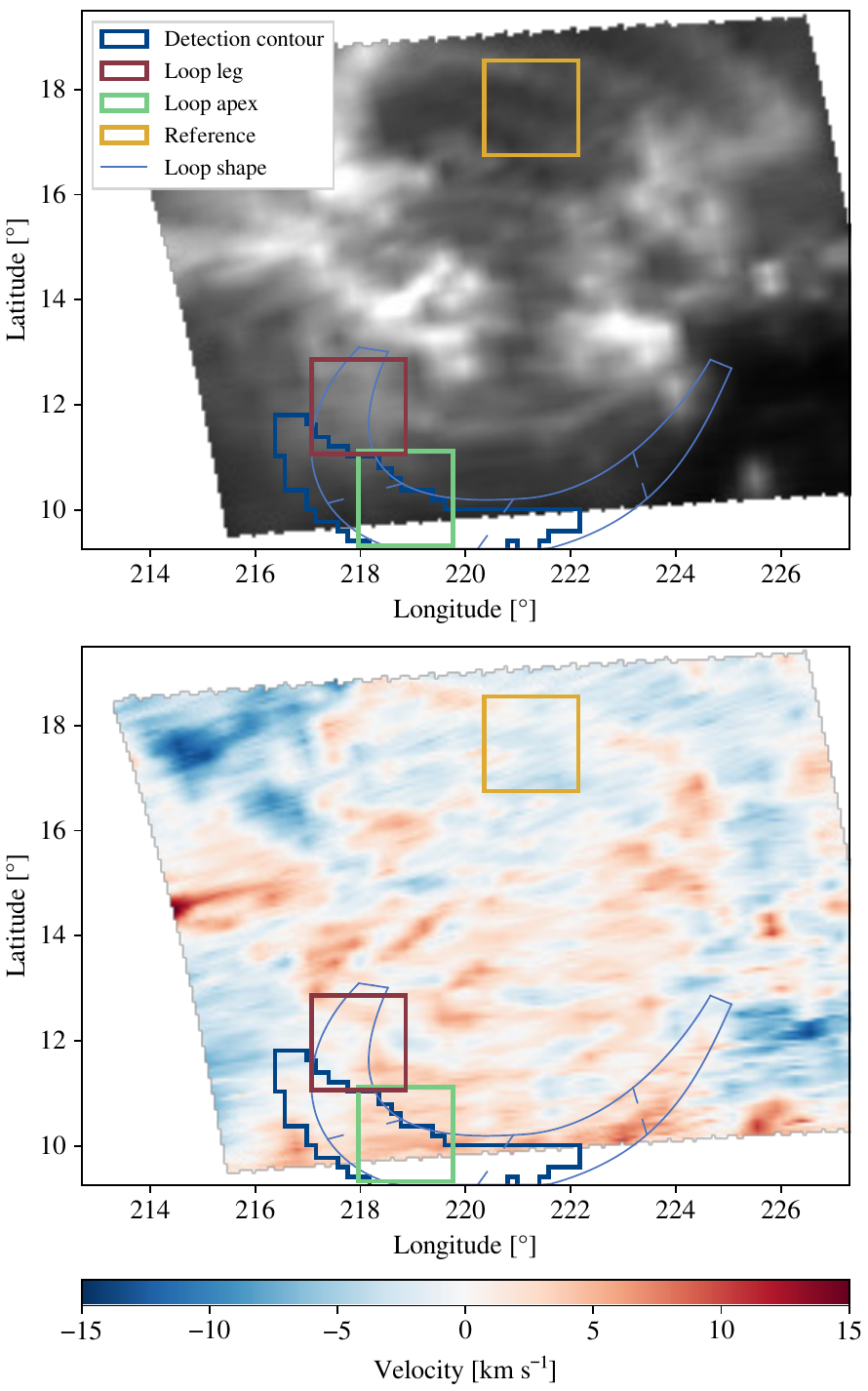}
\caption{
Maps of the \linefexiimain line emission (top) and velocity (bottom) for raster {\FileFont eis\_l0\_20110903\_105615} from dataset~1, projected into Carrington coordinates.
This raster was acquired on \DTMdate{2011-09-03} between \DTMtime{10:56:15} and \DTMtime{11:01:33}.
Several regions of interest are represented on the map:
    the contour in which intensity pulsations are detected with AIA \SI{193}{\angstrom} (blue),
    the contour selected for the loop leg (red),
    the contour selected for the loop apex (green),
    a reference contour outside of the loop (yellow),
    and the shape of the underlying loop (light blue).
The temporal evolution of AIA \SI{193}{\angstrom} can be seen in the online movie.
}
\label{fig:ds1_maps}
\end{figure}

\begin{figure} % fig:ds1_time_series
\includegraphics[width=\columnwidth]{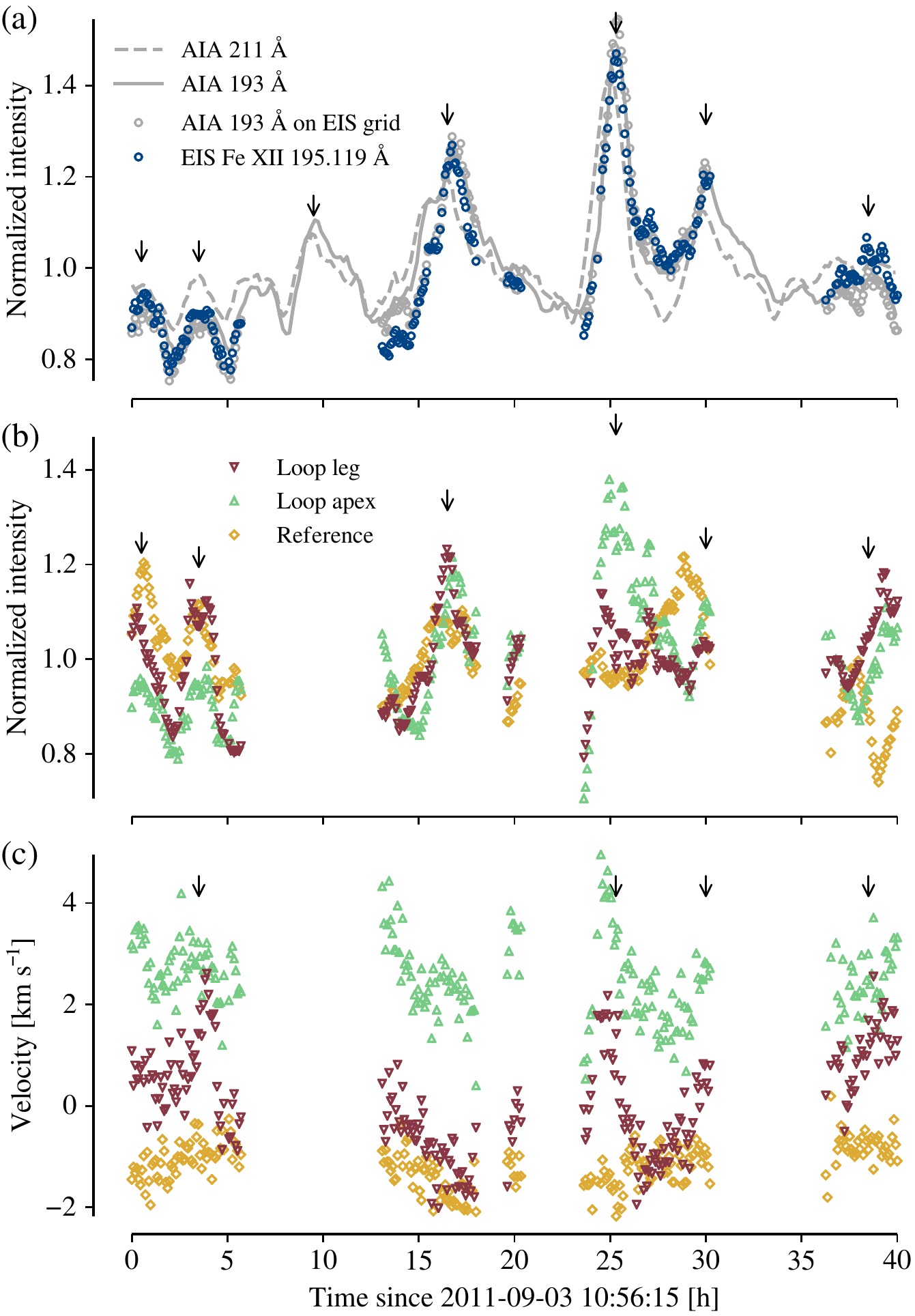}
\caption{
Time series of the intensity and velocity for dataset~1, averaged in the contours defined in \autoref{fig:ds1_maps}.
(a)~Comparison of intensities in the detection contour: EIS \linefexiimain~(\plotDetectionMarker), full resolution AIA \SI{211}{\angstrom}~(\plotAIADetectionLine) and \SI{193}{\angstrom}~(\plotAIAEISLine), and AIA \SI{193}{\angstrom} sampled at the same locations as the EIS rasters~(\plotAIAMarker).
(b)~EIS \linefexiimain intensity normalized to
    (\plotLegMarker~\num{5.6},
     \plotApexMarker~\num{3.9}, and
     \plotReferenceMarker~\num{3.5})
    \si{W.m^{-2}.sr^{-1}}.
(c)~Velocity corrected for the EIS orbital drift.
The arrows mark the positions of the peaks described in the text.
}
\label{fig:ds1_time_series}
\end{figure}

\begin{figure*} % fig:ds1_curvilinear
\centering
\includegraphics[width=\textwidth]{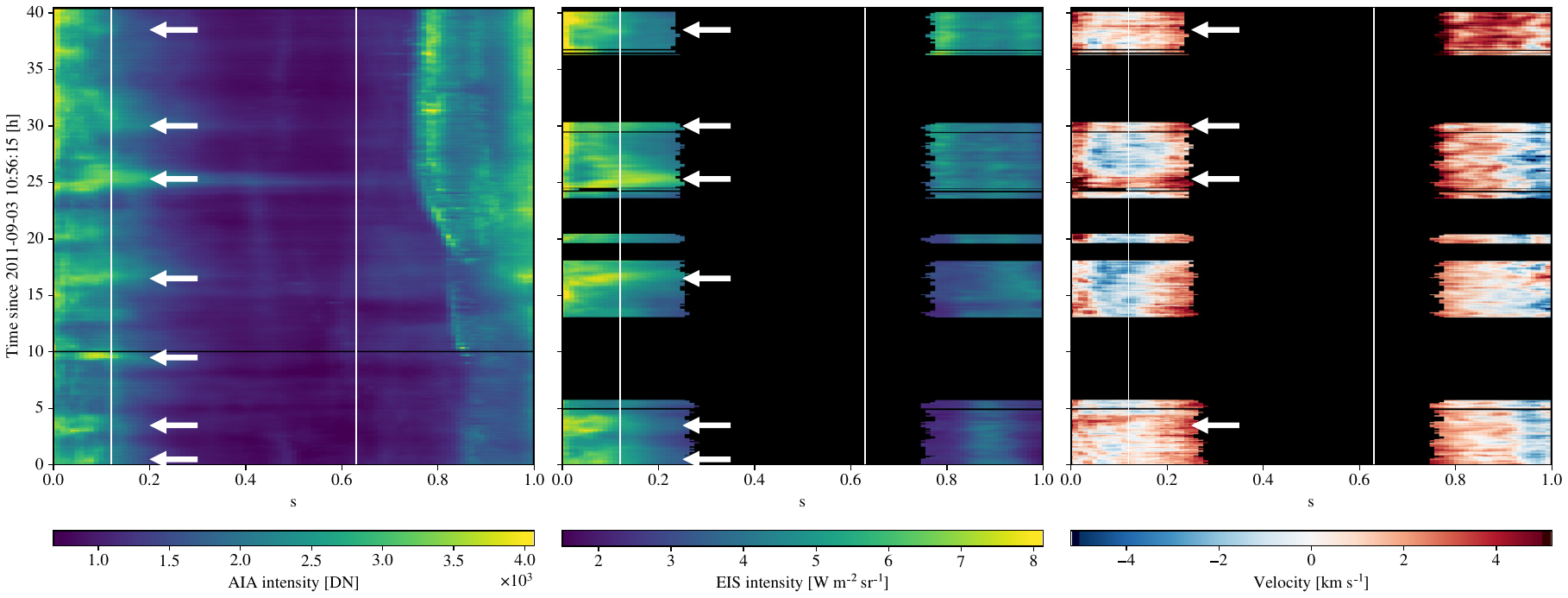}
\caption{
Evolution of the AIA~\SI{193}{\angstrom} intensity (left), EIS \linefexiimain intensity (middle), and velocity (right), as a function of the position along the loop shape of dataset~1 defined in \autoref{fig:ds1_maps}, and of time.
$s = 0$ corresponds to the eastern footpoint, and the values are averaged over the loop width.
Each row of the EIS plots (middle and right) corresponds to a different raster of the dataset.
The vertical lines show the limits of the AIA detection contour, and the arrows correspond to the peaks marked in \autoref{fig:ds1_time_series}.
}
\label{fig:ds1_curvilinear}
\end{figure*}

Dataset~1 corresponds to the observation of active region NOAA~11283, in which intensity pulsations with a period of \SI{5.8}{\hour} were detected in the \SI{211}{\angstrom} channel of AIA, between \DTMdate{2011-09-02}, \DTMtime{13:08:00}, and {\DTMsetdatestyle{AASmd}\DTMdate{2011-09-08}}, \DTMtime{14:18:12}.
The EIS dataset contains 240 rasters recorded between \DTMdate{2011-09-03}, \DTMtime{10:56:15}, and {\DTMsetdatestyle{AASmd}\DTMdate{2011-09-05}}, \DTMtime{02:56:42}, \ie \SI{40.1}{\hour} of observation.
All rasters use the \ang{;;2} slit, \SI{9}{\second} of exposure time, a scan step of \ang{;;6}, and have a FOV of $\ang{;;180}\times\ang{;;152}$.
The FOV is shown in \autoref{fig:ds1_maps}, which contains the intensity and velocity maps from raster {\FileFont eis\_l0\_20110903\_105615} projected into Carrington coordinates corrected for the differential rotation, as well as the region in which the intensity pulsations are detected with AIA.
We select three regions of $\ang{1.8;;}\times\ang{1.8;;}$ in which we examine the evolution of the intensity and velocity: one close to the apex of the loop (green square), one at its eastern leg (red square), and one outside of the pulsating loops (yellow square) that we use as a reference for the velocity.
We verify that the specific shape, position, and size of the regions do not significantly modify the time series.
While some pixels of the regions seem to be outside the loop, they do not reduce the amplitude of the intensity and velocity variations.
The reference region is chosen such that it contains small velocity variations, and no pulsations in the AIA \SI{193}{\angstrom} intensity (we verify that the power spectral density computed using the method presented in \citealp{AuchereEtAl2014} contains no excess power in this region).
These $\ang{1.8;;}\times\ang{1.8;;}$ on the solar sphere correspond to $\ang{;;30}\times\ang{;;30}$ in the plane of the sky for regions at the disk center.
For dataset~1, this corresponds to \SI{5}{pixels} in the solar-X direction and \SI{30}{pixels} in the solar-Y direction, thus \SI{150}{pixels} in each region.
Finally, we add a manually-traced contour that follows the shape of the loop as seen in the AIA \SI{193}{\angstrom} images.

In \autoref{fig:ds1_time_series}, we show the time series associated with dataset~1.
The top panel (\autoref{fig:ds1_time_series}a) shows the intensities from the EIS \linefexiimain line, from AIA \SI{211}{\angstrom} (the channel in which the intensity pulsations are detected), and from AIA \SI{193}{\angstrom} (which is dominated by the \linefexiimain line; \citealp{BoernerEtAl2012}), averaged over the detection contour presented in \autoref{fig:ds1_maps}.
The two AIA channels display similar pulsations, with \SI{193}{\angstrom} peaking after \SI{211}{\angstrom}.
There is a good match between the intensities from EIS \linefexiimain, and AIA \SI{193}{\angstrom}. The small deviations between the two could be caused by other contributions in the passband of AIA, or the fact that the FOV of EIS does not contain the full detection contour (see \autoref{fig:ds1_maps}).
We construct a time series of AIA \SI{193}{\angstrom} intensities sampled at the same locations as the EIS rasters, also shown in \autoref{fig:ds1_time_series}a. This better matches the EIS intensities, thus we conclude that the difference between the EIS and AIA intensities is mainly caused by sampling effects.

The \linefexiimain intensities averaged over the regions shown in \autoref{fig:ds1_maps} are presented in \autoref{fig:ds1_time_series}b.
The time series are divided by their respective average values, which are given in the caption of \autoref{fig:ds1_time_series}.
The intensity in the loop apex and leg have the same behavior as the intensity in the full detection contour.
The intensity in the reference region shows some variations, but these are not always in phase with the variations of the pulsating loop.

The associated velocities are shown in \autoref{fig:ds1_time_series}c.
We estimate the uncertainty on the velocity to \SI{\pm 0.4}{\kmps} by computing the standard-deviation of the time series from the reference region, which contains no feature.
This value is consistent with the usual \SI{\pm 5}{\kmps} uncertainty for one EIS pixel \citep{CulhaneEtAl2007}, divided by the square root of the number of EIS pixels in the region (\SI{150}{pixels} as previously stated), which gives an uncertainty of \SI{0.41}{\kmps}.
% other eg: YoungEtAl2012: 45 s exp time, 1" slit, Fe viii 185.21 Å ±4—5 km s⁻¹

Compared to the reference region, the velocities at the the loop apex (green) and the loop leg (red) show more variance. Some of the fluctuations are in phase with the peaks of intensity.
In particular, four peaks are visible in the loop leg at \num{3.5}, \num{25.3}, \num{30}, and \SI{38.5}{\hour}, which all happen at the same time as intensity peaks. These intensity and velocity peaks are indicated by black arrows on \autoref{fig:ds1_time_series}.
The peaks at \num{3.5} and \SI{25.3}{\hour} have an amplitude of about \SI{3}{\kmps}.
We argue that these are significant because they are above the uncertainty level, and are not present in the reference region.
However, there are no features in velocity associated with the other strong intensity peak around \SI{17}{\hour}.

\autoref{fig:ds1_curvilinear} shows the evolution of the AIA \SI{193}{\angstrom} intensity, the EIS \linefexiimain intensity, and the velocity along the loop shape defined in \autoref{fig:ds1_maps} ($s$ is the position along the loop starting at the eastern footpoint, and the measured parameters are averaged transversely over the loop width).
Each row of the EIS intensity and velocity plots corresponds to a different raster.
In this dataset, there are no data at the loop apex because it is not in the FOV of EIS.
The intensity pulsations are visible along the loop in both AIA and EIS, except near the western footpoint where the emission appears to be dominated by another structure below or above the footpoint.
This structure can be seen in the AIA movie (\autoref{fig:ds1_maps}) and starts expanding near the western footpoint after \DTMdate{2010-11-04} \DTMtime{00:00:00}.
Six intensity peaks (marked by white arrows) are visible with EIS, at \num{0.5}, \num{3.5}, \num{16.5}, \num{25.3}, \num{30}, and \SI{38.5}{\hour}.
Four of these peaks have associated downflows, visible at \num{3.5}, \num{25.3}, \num{30}, and \SI{38.5}{\hour}.

% }}}

% dataset 8 {{{

\begin{figure} % fig:ds8_maps
\centering
\includegraphics[width=\columnwidth]{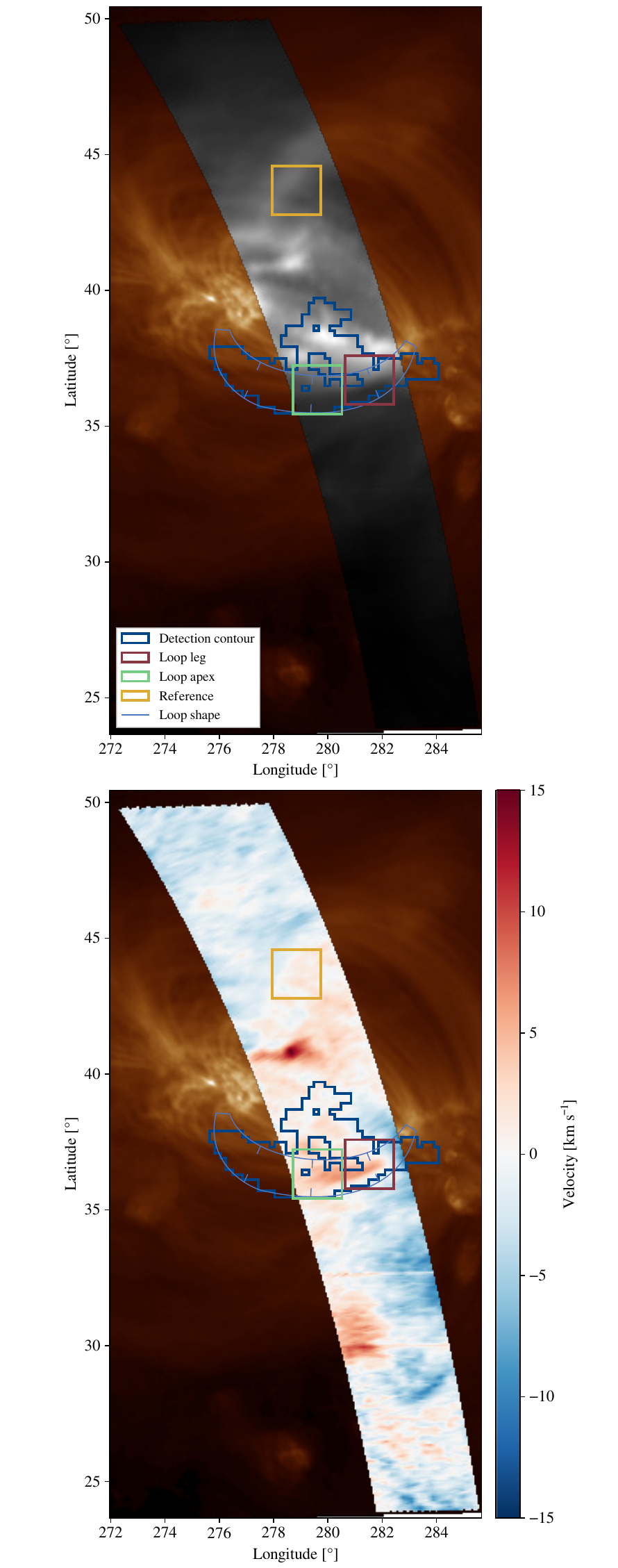}
\caption{
Same as \autoref{fig:ds1_maps}, but with the FOV of raster {\FileFont eis\_l0\_20101103\_211534} from dataset~8,
acquired on \DTMdate{2010-11-03} between \DTMtime{21:15:34} and \DTMtime{21:20:57}.
Both maps are overlaid on a AIA \SI{193}{\angstrom} map to help visualize the structure of the active region.
The temporal evolution of AIA \SI{193}{\angstrom} can be seen in the online movie.
}
\label{fig:ds8_maps}
\end{figure}

\begin{figure} % fig:ds8_time_series
\includegraphics[width=\columnwidth]{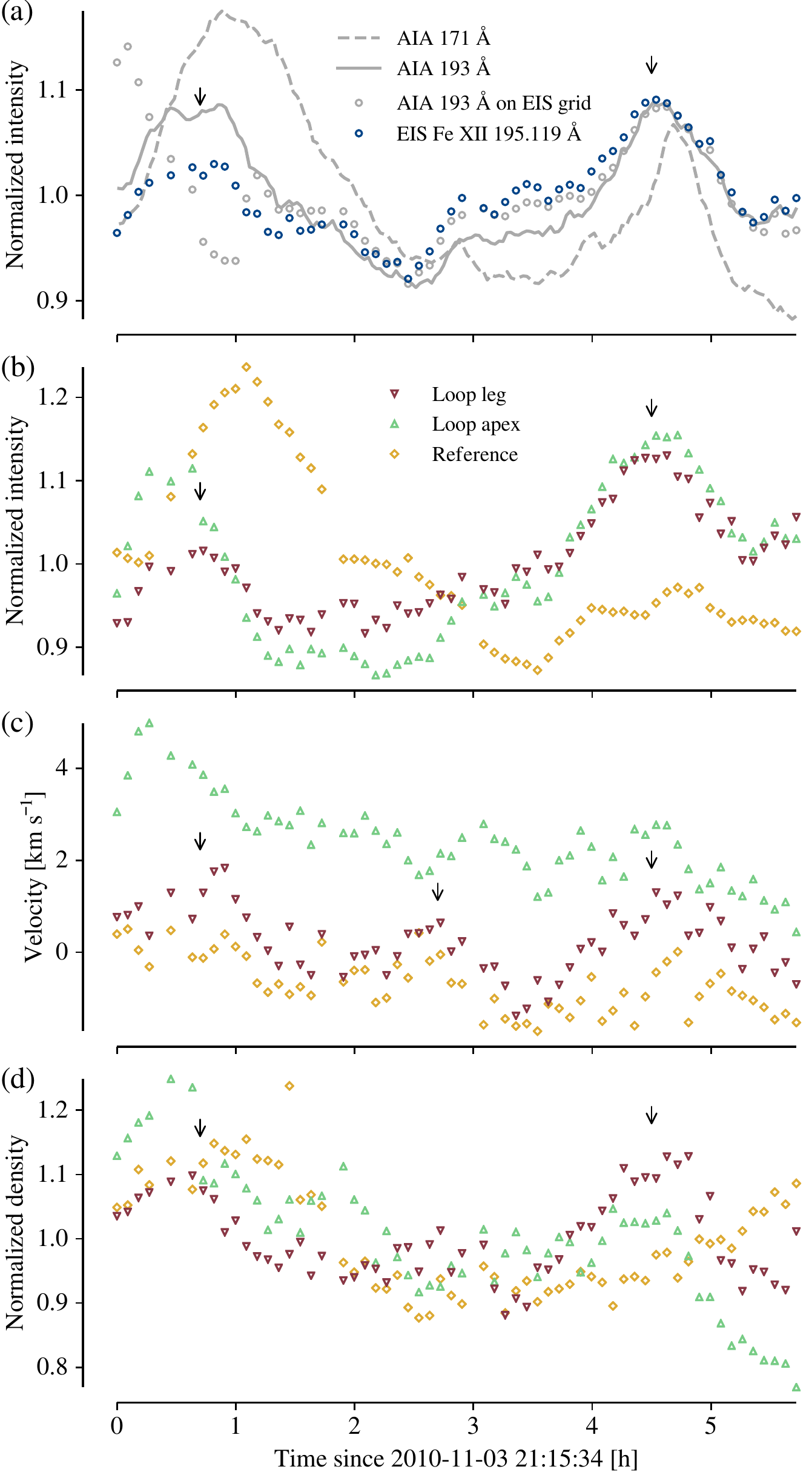}
\caption{
Same as \autoref{fig:ds1_time_series}, but showing the parameters of dataset~8 averaged in the contours defined in \autoref{fig:ds8_maps}.
(a) EIS \linefexiimain and AIA \SI{193}{\angstrom} intensities.
(b) EIS \linefexiimain intensity normalized to
    (\plotLegMarker~\num{6.1},
     \plotApexMarker~\num{3.6}, and
     \plotReferenceMarker~\num{1.0}) \si{W.m^{-2}.sr^{-1}}.
(c) Velocity.
(d) Density normalized to
    (\plotLegMarker~\num{5.0},
     \plotApexMarker~\num{2.1}, and
     \plotReferenceMarker~\num{4.5})$\times\SI{e15}{m^{-3}}$.
}
\label{fig:ds8_time_series}
\end{figure}

\begin{figure*} % fig:ds8_curvilinear
\includegraphics[width=\textwidth]{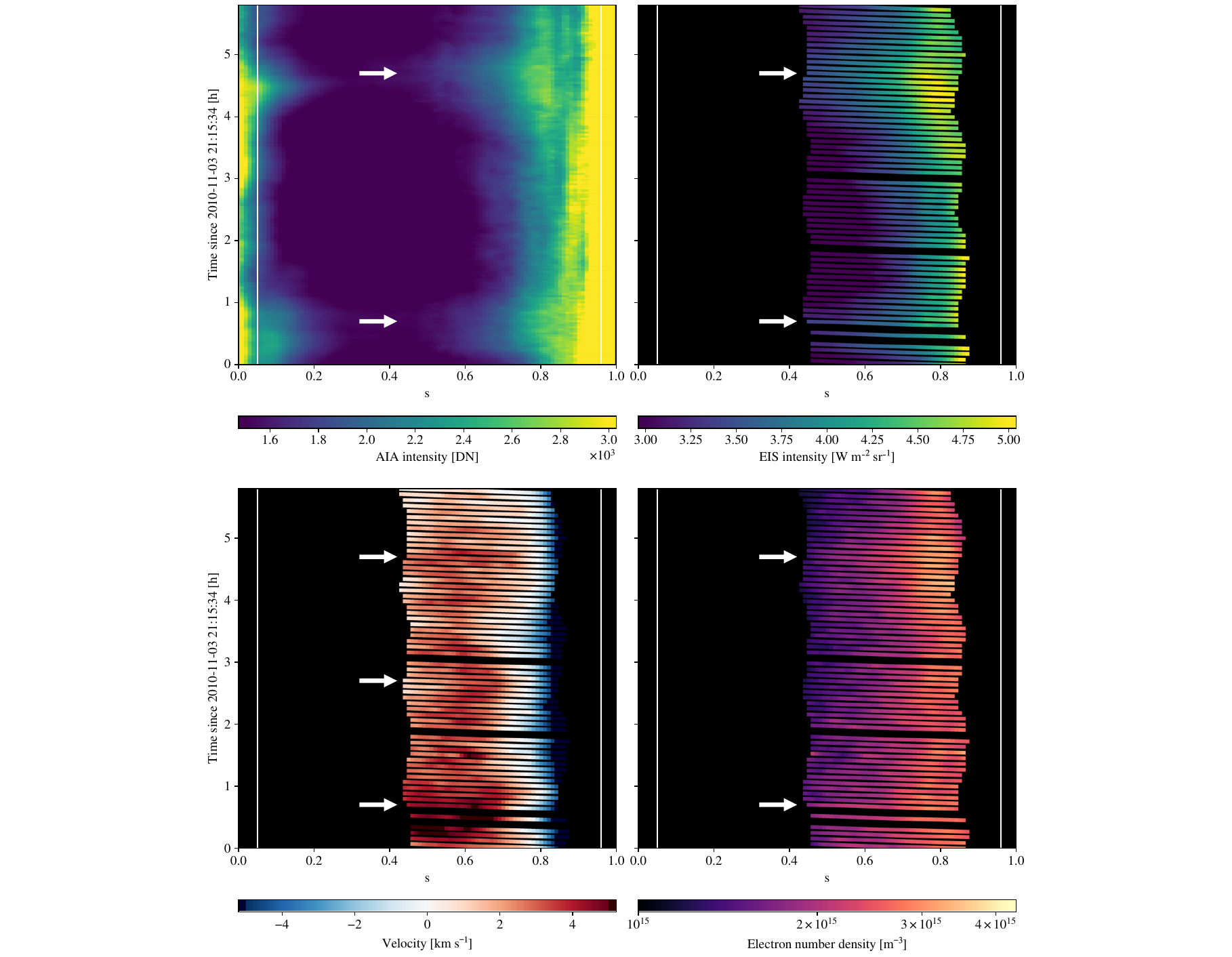}
\caption{
Same as \autoref{fig:ds1_curvilinear}, but for dataset~8, showing the evolution of the AIA \SI{193}{\angstrom} intensity (\textit{top left}), EIS \linefexiimain{} intensity (\textit{top right}), velocity (\textit{bottom left}), and density (\textit{bottom right}) in the loop represented in \autoref{fig:ds8_maps}.
}
\label{fig:ds8_curvilinear}
\end{figure*}

\paragraph*{}\hspace{\parindent}
Dataset~8 corresponds to the observation of NOAA AR~11120, where 3.9-hour intensity pulsations were detected in the \SI{171}{\angstrom} channel of AIA, between \DTMdate{2010-11-02}, \DTMtime{09:10:36}, and {\DTMsetdatestyle{AASmd}\DTMdate{2010-11-08}}, \DTMtime{01:27:12}.
The EIS dataset contains 60 rasters that were recorded between \DTMdate{2010-11-03}, \DTMtime{21:15:35}, and {\DTMsetdatestyle{AASmd}\DTMdate{2010-11-04}}, \DTMtime{02:58:41}. This corresponds to \SI{5.8}{\hour} of observation, which is much shorter than dataset~1 and covers only \num{1.7} pulsation periods.
The rasters use the \ang{;;2} slit, with \SI{20}{\second} of exposure time, a scan step of \ang{;;4}, and have a FOV of $\ang{;;60}\times\ang{;;368}$. Only the western half of the pulsating loop is visible in this narrow FOV.

We perform the same analysis as on the previous dataset, and present similar figures:
\autoref{fig:ds8_maps} shows the FOV of raster {\FileFont eis\_l0\_20101103\_211534}, and the selected regions;
\autoref{fig:ds8_time_series} shows the associated time series;
and \autoref{fig:ds8_curvilinear} shows the evolution of the intensity, velocity, and density along the loop.
Given the \ang{;;4} scan step, each of the $\ang{1.8;;}\times\ang{1.8;;}$ regions shown on \autoref{fig:ds8_maps} corresponds to \SI{225}{pixels} at disk center.
The two channels of AIA show similar intensity pulsations, with \SI{171}{\angstrom} peaking after \SI{193}{\angstrom}. The difference between the AIA \SI{193}{\angstrom} and EIS \linefexiimain intensities appears to be dominated by sampling effects, as in dataset~1.
% The intensity pulsations have a larger amplitude both at the loop apex and in the leg than averaged over the whole detection contour.
The narrow FOV of this dataset makes it difficult to find a reference region in which the intensity does not change much over time, while remaining high enough to allow for accurate velocity measurements. While the region that we select shows some intensity variations (\autoref{fig:ds8_time_series}b), it shows no velocity variations (\autoref{fig:ds8_time_series}c), which is the most important detail to estimate uncertainty on the velocity.

The velocity (\autoref{fig:ds8_time_series}c) has a very small variance in all contours.
However, three small fluctuations are visible in the loop leg, with peaks at \num{0.7}, \num{2.7}, and \SI{4.5}{\hour}.
These are visible in both \autoref{fig:ds8_time_series}c, and \autoref{fig:ds8_curvilinear}, and have an amplitude of less than \SI{2}{\kmps}.
We estimate the uncertainty on the velocity using the values from the reference region to \SI{\pm 0.6}{\kmps}. The observed variations are therefore significant, although very close to the detection limit.
Two of these velocity peaks are associated with intensity maxima and separated by \SI{3.5}{\hour}, \ie approximately one period.
However, the peak at \SI{2.7}{\hour} does not appear to be associated with any intensity feature.

Finally, we measure the density in this dataset, and plot it in \autoref{fig:ds8_time_series}d and \autoref{fig:ds8_curvilinear}.
Similarly to the intensity, the time series are normalized to their average values, which are specified in the caption.
Small density fluctuations ($\sim\SI{20}{\percent}$) are measured in the selected contours, and are visible in \autoref{fig:ds8_time_series}d.
Two maxima are observed in the leg (marked by black arrows), which coincide with the intensity peaks.
A single peak is visible in the density at the apex, which happens about \SI{0.2}{\hour} before the first density peak seen in the loop leg.
The density peak at the apex is accompanied by co-temporal intensity and velocity peaks at the apex, visible in \autoref{fig:ds8_time_series} b and c.
The density variations are also visible in \autoref{fig:ds8_curvilinear}, which shows the evolution of the density along the loop.
The density in the other regions is not correlated with the variations in intensity.

% }}}

\subsection{Datasets with no pulsations because of instrumental limitations}\label{subsec:no_pulsations}

% dataset 2 - No pulsations because of the SNR {{{

\begin{figure} % fig:ds2_maps
\includegraphics[width=\columnwidth]{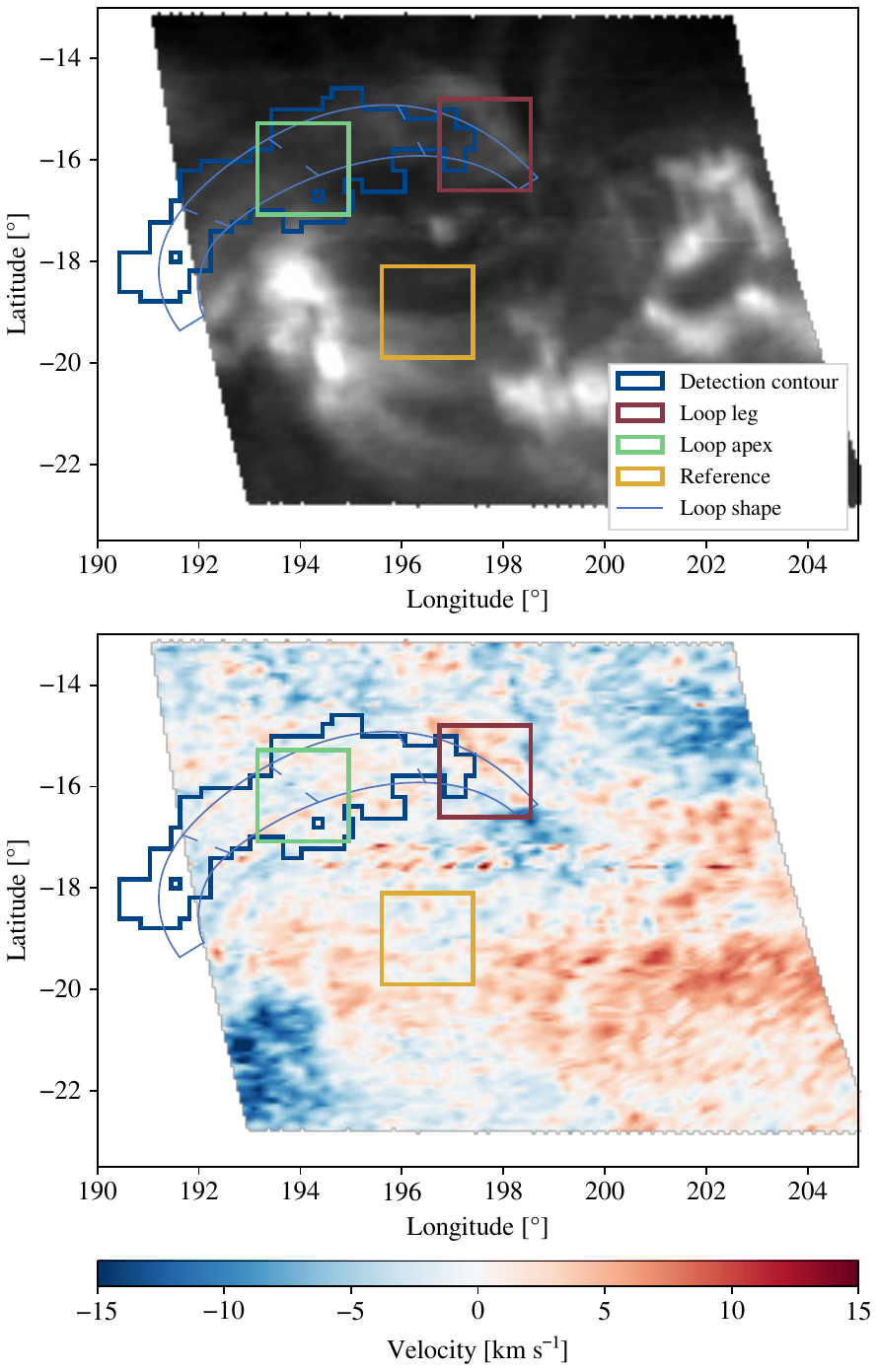}
\caption{
Same as \autoref{fig:ds1_maps}, but with the FOV of raster {\FileFont eis\_l0\_20120608\_230140} from dataset~2,
acquired on \DTMdate{2012-06-08} between \DTMtime{23:01:40} and \DTMtime{23:05:51}.
The temporal evolution of AIA \SI{193}{\angstrom} can be seen in the online movie.
}
\label{fig:ds2_maps}
\end{figure}

\begin{figure} % fig:ds2_time_series
\includegraphics[width=\columnwidth]{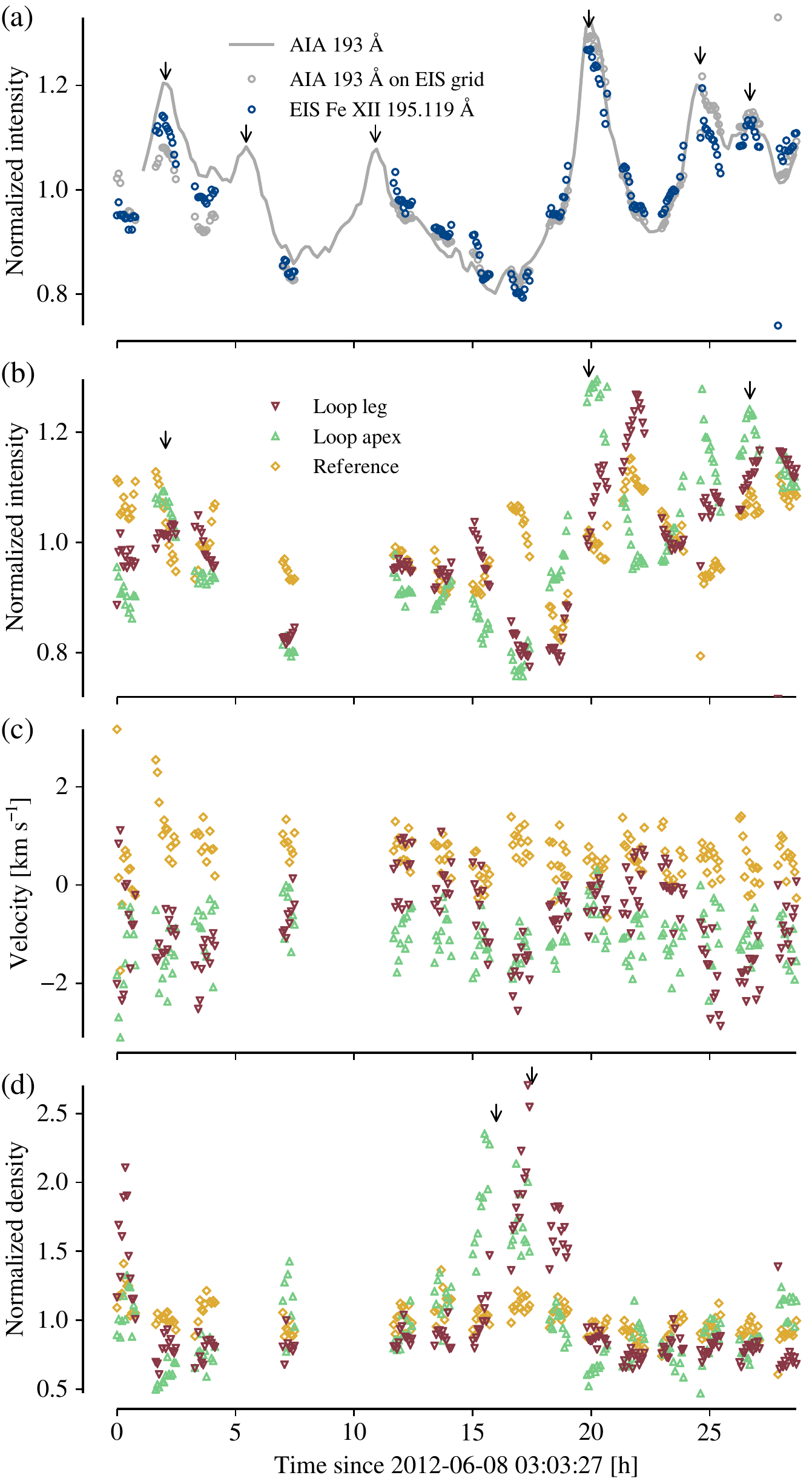}
\caption{
Same as \autoref{fig:ds1_time_series}, but showing the parameters of dataset~2 averaged in the contours defined in \autoref{fig:ds2_maps}.
(a) EIS \linefexiimain and AIA \SI{193}{\angstrom} intensities.
(b) EIS \linefexiimain intensity normalized to
    (\plotLegMarker~\num{2.9},
     \plotApexMarker~\num{2.2}, and
     \plotReferenceMarker~\num{3.2}) \si{W.m^{-2}.sr^{-1}}.
(c) Velocity.
(d) Density normalized to
    (\plotLegMarker~\num{2.1},
     \plotApexMarker~\num{2.8}, and
     \plotReferenceMarker~\num{2.8})$\times\SI{e15}{m^{-3}}$.
}
\label{fig:ds2_time_series}
\end{figure}

\begin{figure*} % fig:ds2_curvilinear
\includegraphics[width=\textwidth]{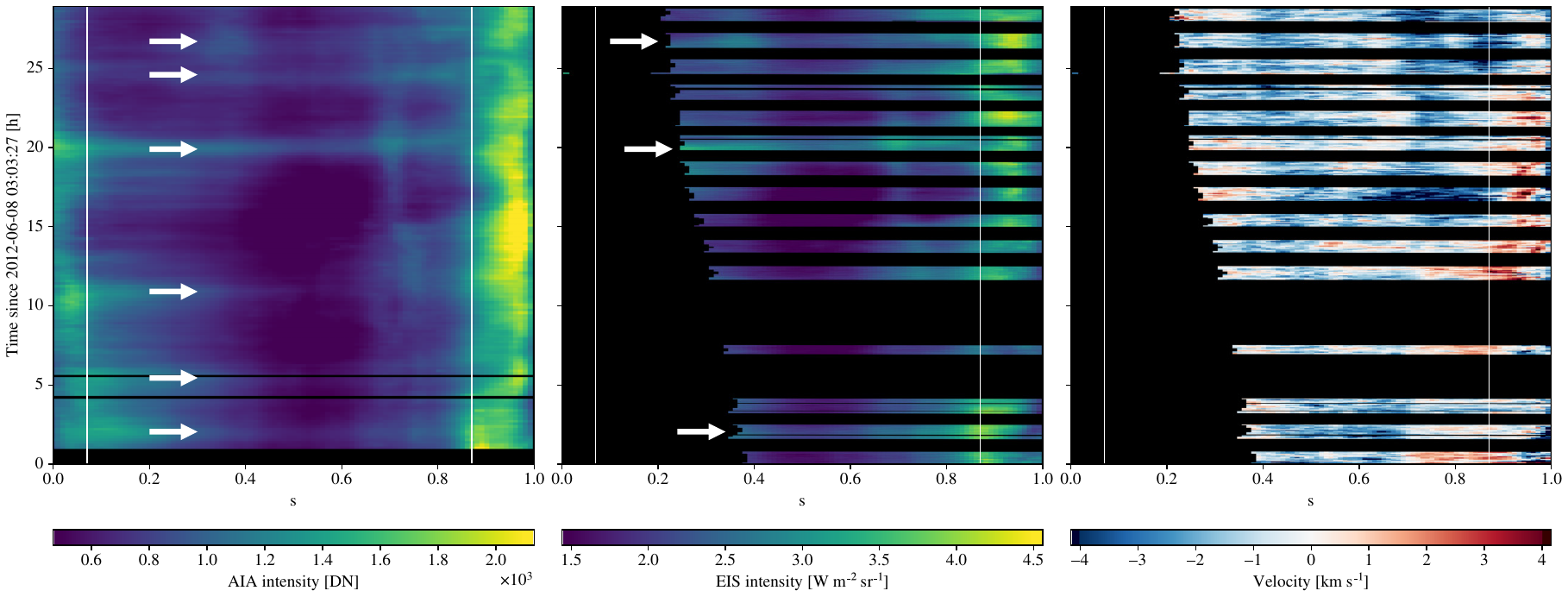}
\caption{
Same as \autoref{fig:ds1_curvilinear}, but for dataset~2, showing the evolution of the intensity and velocity in the loop represented in \autoref{fig:ds2_maps}.
}
\label{fig:ds2_curvilinear}
\end{figure*}

Dataset~2 corresponds to the observation of NOAA AR~11494, in which pulsations were detected in the \num{193} and \SI{335}{\angstrom} channels of AIA between \DTMdate{2012-06-03} \DTMtime{18:44:31}, and {\DTMsetdatestyle{AASmd}\DTMdate{2012-06-09}} \DTMtime{11:39:31}, with a period of \SI{4.9}{\hour}.
The EIS dataset contains 178 rasters recorded between \DTMdate{2012-06-08} \DTMtime{03:03:27}, and {\DTMsetdatestyle{AASmd}\DTMdate{2012-06-09}} \DTMtime{07:42:53}, \ie \SI{28.7}{\hour} of continuous observation.
All rasters use the \ang{;;2} slit, a \SI{3}{\second} exposure time, a scan step of \ang{;;3}, and have a FOV of $\ang{;;152}\times\ang{;;162}$.
At this exposure time, the SNR is 3~times lower than in dataset~1, and 7~times than in dataset~8.

We present similar figures as for the previous dataset: the FOV of raster {\FileFont eis\_l0\_20120608\_230140} and the regions of interest are shown on \autoref{fig:ds2_maps} (each region corresponds to \SI{300}{pixels} at disk center). The associated time series are shown on \autoref{fig:ds2_time_series}, and the evolution of the intensity and velocity along the loop on \autoref{fig:ds2_curvilinear}.
For this dataset, the intensity in the loop leg seems to peak after the intensity at the apex (\autoref{fig:ds2_time_series}b). This is consistent with falling material at coronal temperatures, but such behaviour is not observed in the other datasets.
Contrary to datasets~1 and 8, there is no significant evolution of the velocity, which is most likely explained by the fact that the SNR is significantly lower in this dataset.
Velocity variations can be seen in the western part of the loop in \autoref{fig:ds2_curvilinear}, but these are not correlated with the intensity peaks.
However, a prominent peak is visible in the density around \SI{16}{\hour} at the apex, and around \SI{17.5}{\hour} in the loop leg (\autoref{fig:ds2_time_series}d).
This falls between two intensity peaks at \num{11} and \SI{20}{\hour}.

% }}}

% dataset 11 - No pulsations because of the cadence {{{

\begin{figure} % fig:ds11_maps
\centering
\includegraphics[width=\columnwidth]{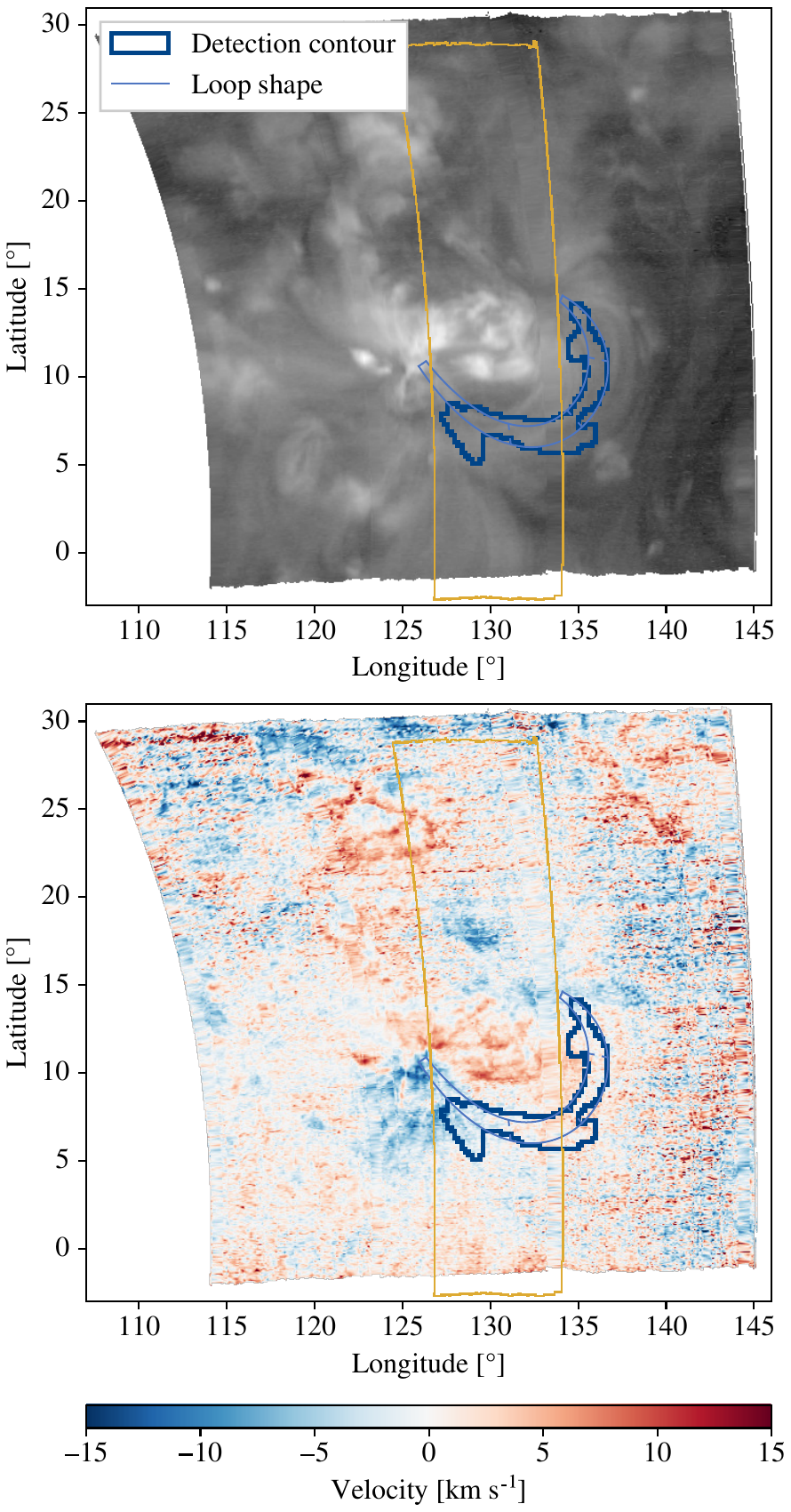}
\caption{
Same as \autoref{fig:ds1_maps}, but with the FOV of raster {\FileFont eis\_l0\_20140810\_042212} from dataset~11,
acquired on \DTMdate{2014-08-10} between \DTMtime{04:22:12} and \DTMtime{05:29:51}.
The AIA detection contour is shown in dark blue, the loop shape in light blue, and the FOV of raster {\FileFont eis\_l0\_20140810\_192924} is shown in yellow.
The temporal evolution of AIA \SI{193}{\angstrom} can be seen in the online movie.
}
\label{fig:ds11_maps}
\end{figure}

\begin{figure} % fig:ds11_time_series
\includegraphics[width=\columnwidth]{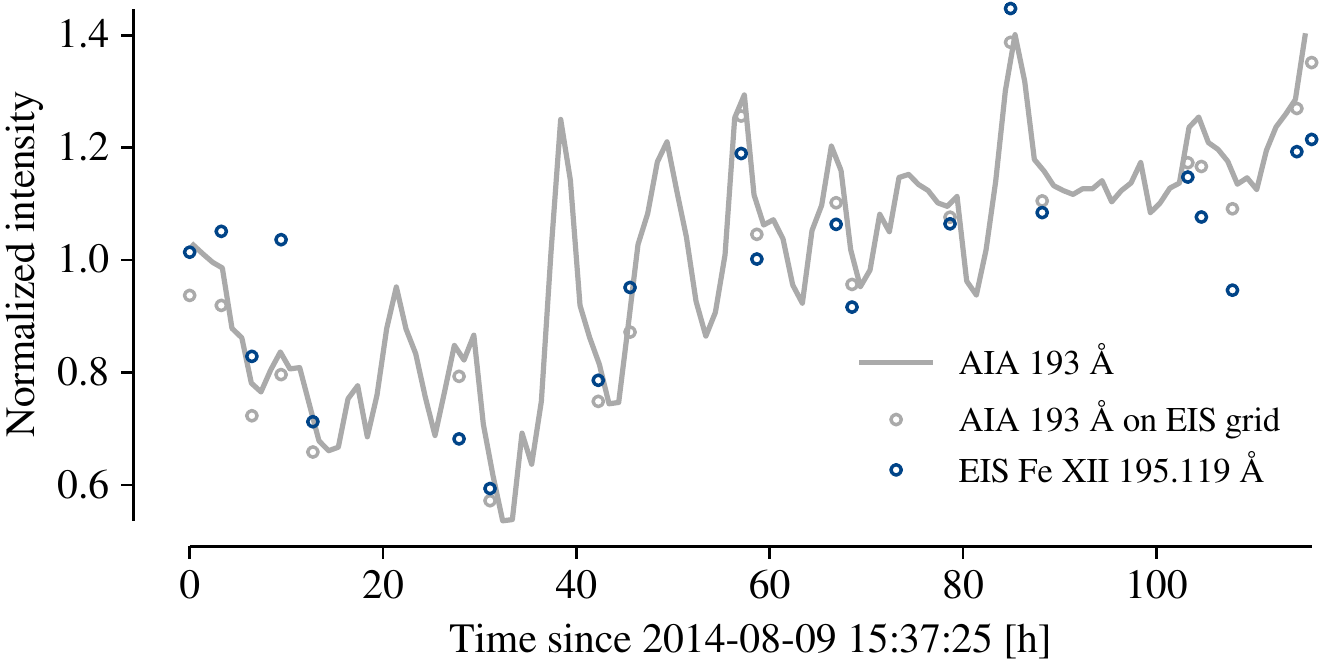}
\caption{
Intensity evolution for dataset~11, averaged in the detection contour shown in \autoref{fig:ds11_maps}:
AIA \SI{193}{\angstrom} at full resolution (\plotAIAEISLine), EIS \linefexiimain intensity (\plotDetectionMarker), and AIA \SI{193}{\angstrom} sampled at the same locations as the 21 EIS rasters (\plotAIAMarker).
}
\label{fig:ds11_time_series}
\end{figure}

\begin{figure*} % fig:ds11_curvilinear
\includegraphics[width=\textwidth]{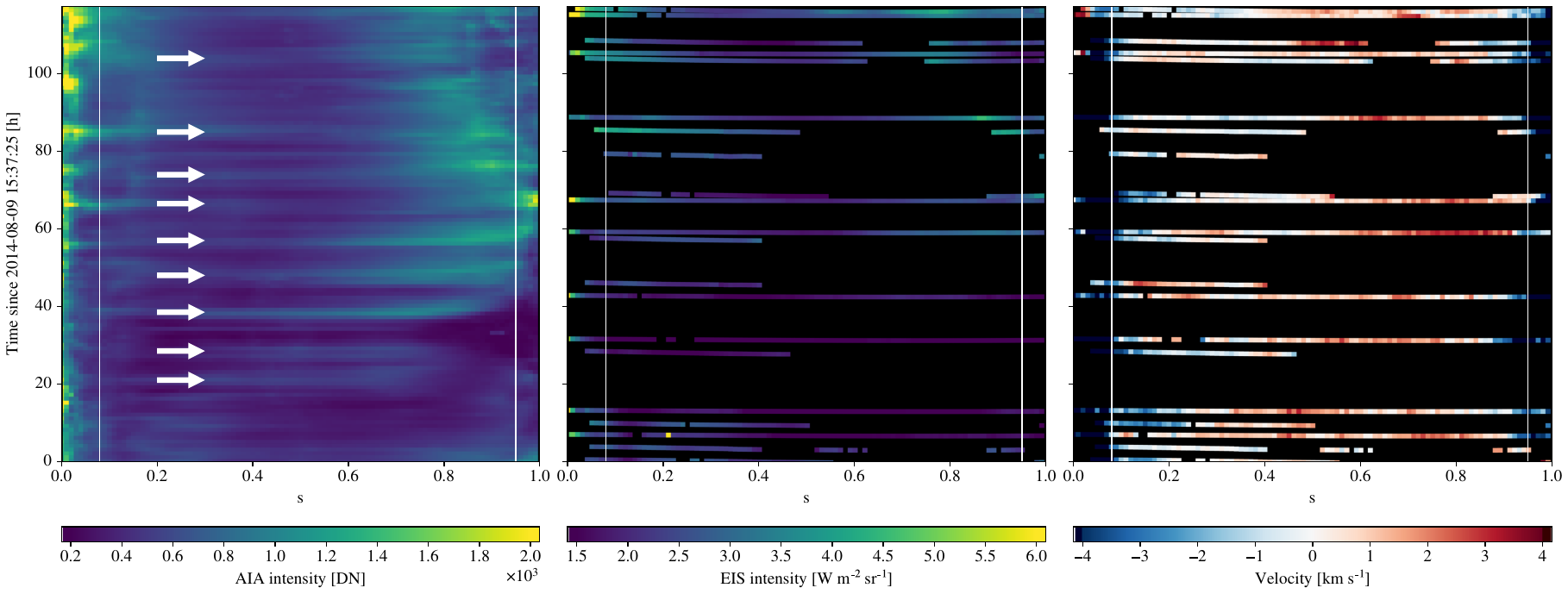}
\caption{
Same as \autoref{fig:ds1_curvilinear}, but for dataset~11, showing the evolution of the intensity and velocity in the loop represented in \autoref{fig:ds11_maps}.
}
\label{fig:ds11_curvilinear}
\end{figure*}

\paragraph*{}\hspace{\parindent}
Dataset~11 corresponds to the observation of AR NOAA~12135 in which pulsations were detected in the \SI{193}{\angstrom} channel of AIA, between \DTMdate{2014-08-09} \DTMtime{10:56:06}, and {\DTMsetdatestyle{AASmd}\DTMdate{2014-08-15}} \DTMtime{12:26:18}, with a period of \SI{5.8}{\hour}.
The dataset contains 21 rasters acquired between \DTMdate{2014-08-09} \DTMtime{15:37:25}, and {\DTMsetdatestyle{AASmd}\DTMdate{2014-08-14}} \DTMtime{11:41:13}, \ie \SI{116.8}{\hour} of observation with a very low cadence of one raster every \SI{5.6}{\hour} on average.
It is composed of two kinds of rasters:
the first uses study {\FileFont HPW022\_VEL\_480x512v1} (ID: \href{http://solarb.mssl.ucl.ac.uk:8080/SolarB/ShowEisStudy.jsp?study=480}{480}), with an exposure time of \SI{15}{\second}, and a wide FOV of $\ang{;;480}\times\ang{;;512}$;
the second kind uses the EIS study {\FileFont HPW021\_VEL\_120x512v2} (ID: \href{http://solarb.mssl.ucl.ac.uk:8080/SolarB/ShowEisStudy.jsp?study=428}{428}), with an exposure time of \SI{45}{\second}, and a relatively narrow FOV of $\ang{;;120}\times\ang{;;512}$.
All rasters use the \ang{;;1} slit.

\autoref{fig:ds11_maps} shows the intensity and velocity maps of raster {\FileFont eis\_l0\_20140810\_042212} (wide FOV), the contour of raster {\FileFont eis\_l0\_20140810\_192924} (narrow FOV), the region in which the intensity pulsations are detected in AIA, and a loop shape that extends this detection contour towards the footpoints. The eastern part of the loop is covered by both raster types, while the western part is only seen in rasters with a wide FOV.
\autoref{fig:ds11_time_series} shows the evolution of the intensity in the detection contour. The \num{5.8}-hour pulsations are clearly visible in the AIA \SI{193}{\angstrom} time series, and the EIS \linefexiimain intensity matches its evolution. However, the cadence of the EIS rasters is not high enough to detect the pulsations with these data only.
\autoref{fig:ds11_curvilinear} shows the evolution of the intensity and velocity along the loop contour defined in \autoref{fig:ds11_maps}.
Despite a good SNR that allows for accurate velocity measurements, no velocity nor intensity pulsations can be seen in the EIS plots.
However, downflows are detected in the western part of the loop ($0.6 < s < 0.9$), while the eastern legs contains either upflows or no velocities.
This is compatible with either a static flow along the loop, or the expected pulsations, but the cadence does not allow us to discriminate between the two scenarios.

% }}}

\section{Discussion} \label{sec:discussion}

\subsection{Magnitude of the measured downflows}
\label{subsec:discussion_amplitude}

\begin{figure} % fig:los_integration
\includegraphics[width=\columnwidth]{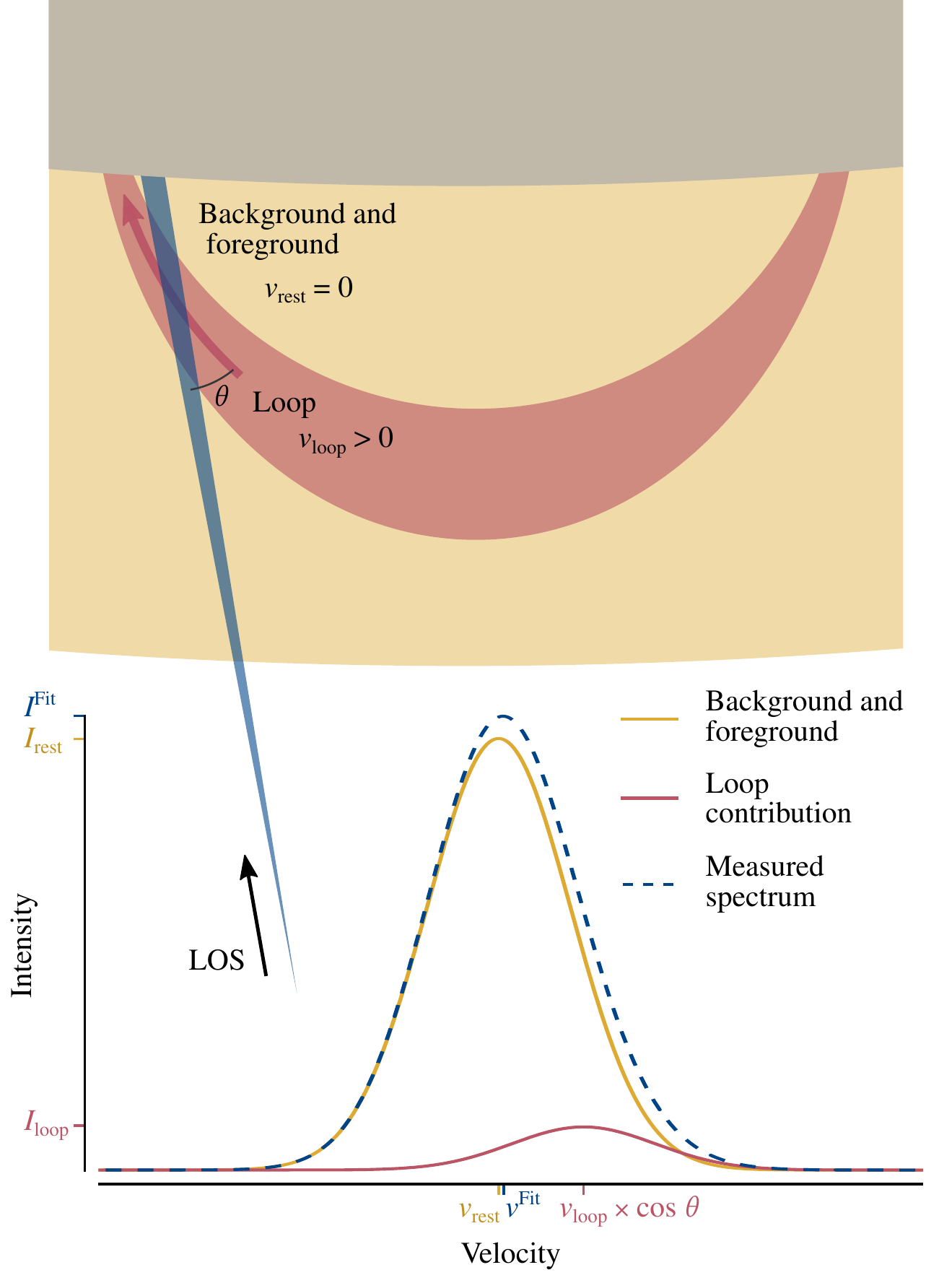}
\caption{
Sketch representing the integration along the LOS of a spectral line emitted by plasma flowing in a loop (red), and the contribution from static plasma in the background and foreground (yellow).
Even if the component from the loop has an important spectral shift, the resulting line (dashed blue) is close to the rest wavelength, because the loop accounts only for \SI{10}{\percent} to \SI{30}{\percent} of the total emission.
}
\label{fig:los_integration}
\end{figure}

The one-dimensional hydrodynamic simulations of loops that reproduce the observed long-period intensity pulsations also predict periodic plasma flows, where the velocity along the loop changes over one cycle with an amplitude of about \SI{60}{\kmps} \citep{MikicEtAl2013, FromentEtAl2017, FromentEtAl2018}.
Such variations in velocity should be easy to detect with \textit{Hinode}/EIS, which can measure velocities with a precision of about \SI{5}{\kmps} in a single pixel \citep{CulhaneEtAl2007}.
We searched for such pulsations by analyzing 11 EIS datasets, 9 of which had sufficient cadence to allow for the detection of pulsations.
Yet, we did not detect velocity pulsations with the expected amplitude in any of these datasets.
Instead of this, we detect velocity variations with an amplitude of \num{2} to \SI{4}{\kmps} in two datasets, which are the ones that have the highest SNR, and the most contrasted intensity variations.
The measured velocity variations are therefore lower by at least a factor of \num{10} than those produced in the simulations.

This apparent discrepancy is caused by the fact that coronal loops are only \SI{10}{\percent} to \SI{30}{\percent} brighter than the background when observed in EUV \citep{DelZannaMason2003, AschwandenNightingale2005, AschwandenEtAl2008}, and therefore only contribute to a small fraction of the emission integrated over the full line of sight (LOS).
Let’s consider a LOS filled with plasma at rest, that intersects at an unknown angle $\theta$ with a single loop inside which plasma flows at a velocity $v_\text{loop}$, as illustrated in \autoref{fig:los_integration}.
In this situation, the velocity projected on the LOS is $v_\text{loop} \cos\theta$, and a given electronic transition with an energy equivalent to the rest wavelength $\lambda_0$ (\linefexiimain in our case) would result in two distinct contributions: a bright contribution centered on $\lambda_0$ emitted by everything outside of the loop, and a dimmer contribution emitted by the plasma flowing in the loop, centered on $(v_\text{loop} \cos\theta/c + 1) \lambda_0$, and only \SI{10}{\percent} to \SI{30}{\percent} as bright as the first line.
Because of the combined Doppler broadening and instrumental width, line profiles observed by EIS have typical full widths at half-maximum (FWHM) of \SI{60}{\milli\angstrom}, or \SI{95}{\kmps} \citep{KorendykeEtAl2006, BrownEtAl2008}.
With an expected separation of $\cos\theta\times \SI{60}{\kmps}$, the two lines are therefore blended, and retrieving the velocity of the fainter component is not straightforward.
This may be achieved by fitting the two lines with either a single, or two Gaussian profiles.

We tested these two approaches by performing Monte-Carlo simulations, in which we generate synthetic spectra similar to the one described above (\ie two gaussian profiles at positions $v_\text{rest} = 0$ and $v_\text{loop}\cos\theta > 0$, intensities $I_\text{rest} > I_\text{loop}$, and a common FWHM $\Delta$), add photon noise, and fit the spectra with either one or two Gaussian profiles.
By repeating this operation a large number of times for different realizations of the noise, we can estimate the probability of correctly retrieving the input parameters.
We explore different values for the wavelength separation, intensity ratio, and SNR of the two lines.
This is detailed in \autoref{sec:mc_fit}.
We draw two conclusions from these simulations:
1. given the SNR of the EIS observations, the velocity of the second component cannot be estimated with a two-gaussians fit, because the locations of the two fitting Gaussians are decorrelated from the input for $v_\text{loop}$ lower than \SI{80}{\kmps} or \SI{150}{\kmps} depending on the SNR (\autoref{fig:mc_dgb});
2. when performing a single-gaussian fit, the retrieved velocity $v^\text{Fit}$ is systematically lower than the one in the loop (\autoref{fig:mc_ratio}), with:

\begin{equation}\label{eqn:vfit}
v^\text{Fit} \lesssim v_\text{loop} \cos\theta \times I_\text{loop} / I_\text{rest}, \ \text{when}\  |v_\text{loop} \cos\theta| < \Delta / 2.
\end{equation}

These simulations justify the use of a single-gaussian fit to retrieve the velocity of a faint component with a small wavelength separation, and provide a new way to interpret the fit results.
Double gaussian fitting is therefore more suited to larger separations (see, \eg \citealp{ImadaEtAl2008, DollaZhukov2011} who applied this method to retrieve separations of \num{50}--\SI{100}{\kmps} from EIS spectra), while the B--R asymmetry index \citep{DePontieuEtAl2009} is adapted to more complex line profiles, but do not allow for straightforward velocity measurements.

We use the intensity contrast presented in \autoref{tab:datasets} as an estimation of $I_\text{loop} / I_\text{rest}$ to compute a lower bound to the amplitude of the velocity variations in the loops using the above \autoref{eqn:vfit}: $v_\text{loop, min} \cos\theta = v^\text{Fit} \times I_\text{rest}/I_\text{loop}$.
In dataset~1, we measured variations of \SI{3 \pm .4}{\kmps} and an intensity contrast of \SI{50}{\percent}, which translates to $v_\text{loop, min} \cos\theta = \SI{6}{\kmps}$.
For dataset~8, the measured variations are of \SI{2 \pm .6}{\kmps} with a contrast of \SI{20}{\percent}, which gives $v_\text{loop, min} \cos\theta = \SI{10}{\kmps}$.
These values are closer, although still lower, to those produced in the simulations.
Part of this difference results from the projection of the LOS.
In the case of dataset~1, the measured velocity could be further reduced by the orbital drift correction, for which we used the velocity averaged over the FOV (\autoref{sec:method}). The reference region used to correct the orbital drift therefore includes the pulsating loops, which could slightly attenuate the velocity variations.
The presence of counter-streaming flows \citep{FangEtAl2013, FangEtAl2015, XiaEtAl2017} may further explain the small velocity variations. Such flows would indeed add a blueshifted contribution to the spectral line, which would shift its centroid towards lower velocities. However, the current analysis does not allow to tell whether such flows are present in the loops.

Finally, the LOS integration effect can also explain why pulsations are not seen in all datasets: in most datasets, the measured velocity would be reduced by the background and foreground to the point that it falls below the detection threshold of EIS.
Datasets~1 and 8 where velocity variations are measured are those with the most favorable combination of SNR and intensity contrast.

\subsection{Time shifts between the intensity, the velocity, and the density}
\label{subsec:discussion_phase}

\begin{figure} % fig:froment+2017_time_series
\includegraphics[width=\columnwidth]{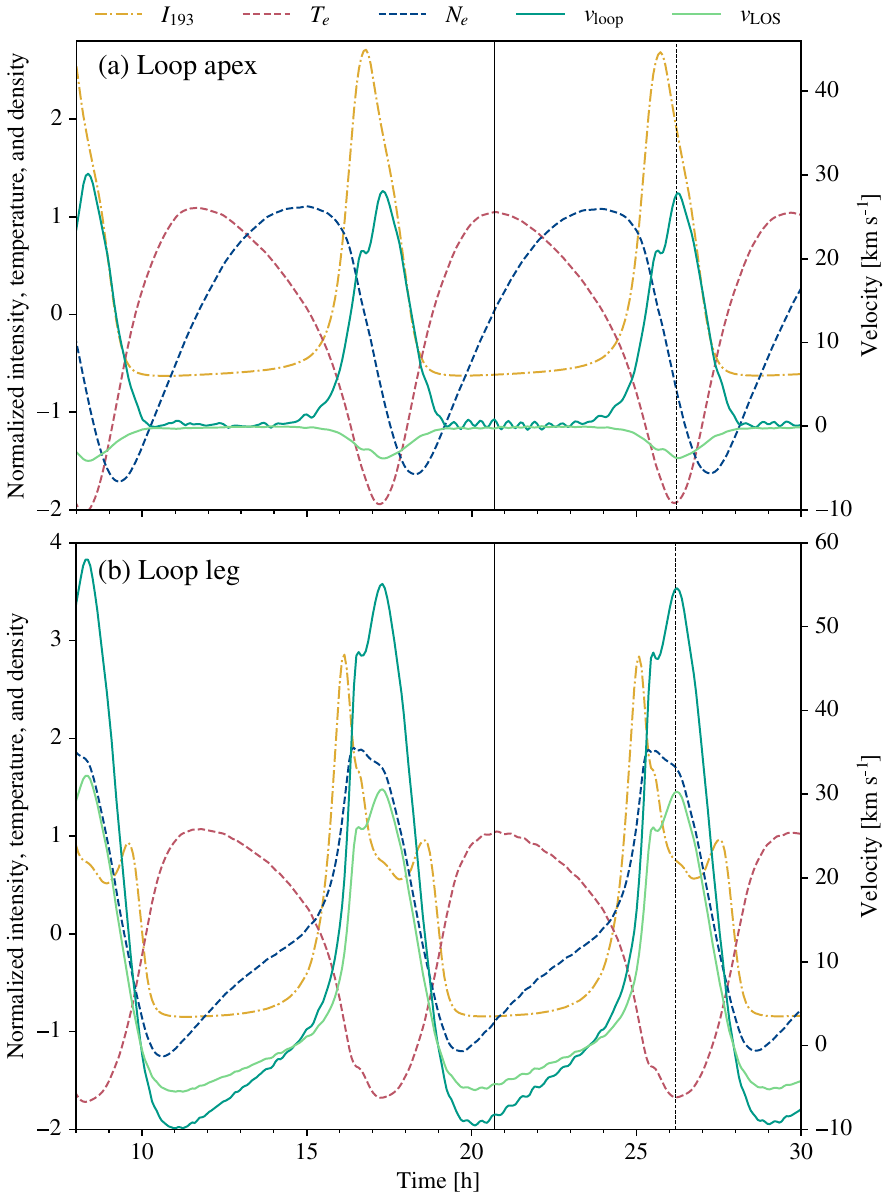}
\caption{
Synthetic AIA \SI{193}{\angstrom} intensity $I_{193}$, temperature $T_e$, density $N_e$, and velocity ($v_\text{loop}$ along the loop, and $v_\text{LOS}$ projected along the line of sight), in the loop simulated by \citet{FromentEtAl2017}, and adapted from Figure 9 of their paper.
% which shows the synthetic intensities in the 6 coronal channel of AIA, as well as the temperature and density, averaged at the loop apex.
Top: values near the apex, averaged between \num{170} and \SI{255}{\mega\meter} of the \SI{367}{Mm}-long-loop.
Bottom: values in the loop leg, averaged between \num{320} and \SI{340}{\mega\meter}.
The intensity, temperature, and density are divided by their standard deviation and shifted by their average.
$N_e$ peaks before $I_{193}$, $v_\text{loop}$, and $v_\text{LOS}$ at the apex, but all peak roughly at the same time in the leg of the loop.
}
\label{fig:froment+2017_time_series}
\end{figure}

We now investigate how the intensity, velocity, and density signals are shifted relatively to each other, as it is a signature of TNE.
In datasets~1 and 8, almost all observed velocity peaks happen at the same time as the \linefexiimain intensity peaks.
The density peaks are less consistent: it appears to be in phase with the intensity in dataset~8, while it is in opposition of phase in dataset~2.

In order to better understand these behaviors, we take a new look at simulation results from \citet{FromentEtAl2017}, which reproduced the intensity pulsations observed with AIA for one of the events presented by \citep{FromentEtAl2015}.
This simulation was performed by \citet{FromentEtAl2017}, who used the method described by \citet{MikicEtAl2013} to computes the hydrodynamic evolution of the plasma along a fixed magnetic field line with a non-uniform area expansion.
Although this simulation was performed for a different event than the ones presented in the current study, we use it to get a global idea of the evolution of a loop undergoing TNE cycles.
In \autoref{fig:froment+2017_time_series} \citep[adapted from Figure 9 of][]{FromentEtAl2017}, we present the evolution of the AIA \SI{193}{\angstrom} intensity, temperature, density, and velocity (along the loop and projected along the LOS of AIA or EIS), averaged at the loop apex (top) and in the western leg of the loop (bottom), which is the leg towards which the condensations fall.
We first note that the velocity in the western leg peaks to \SI{55}{\kmps} along the loop, which corresponds to only \SI{30}{\kmps} when projected along the LOS of AIA or EIS, as mentioned in \autoref{sec:intro}.
In the eastern leg (not shown in \autoref{fig:froment+2017_time_series}), the velocity only reaches \SI{15}{\kmps} along the loop, which is even smaller than the velocity at the apex. This is not surprising given that the condensations do not flow towards this leg.
At the apex, the density peaks before the intensity, while the velocity peaks roughly at the same time as the intensity. In the loop leg, however, all parameters peak approximately at the same time.
Therefore we expect the velocity to be in phase with the intensity everywhere in the loop, but the density should peak at the loop apex before peaking in the leg.

In dataset~8, two density peaks occur in the leg at the same time as the two intensity peaks, around \num{0.7} and \SI{4.5}{h} (\autoref{fig:ds8_time_series}).
One density peak is visible at the apex at \SI{0.5}{h}, just before the first peak that occurs in the leg at \SI{0.7}{h}.
The time shifts of these three density peaks is consistent with the simulations.
However, the second density peak is not visible at the apex (this could be due to no variations or insufficient contrast), and we cannot fully test the fact that the density should peak before the intensity at this location.
A single but prominent density peak is detected in dataset~2, which arises first at the apex at \SI{16}{h}, and then in the leg at \SI{17.5}{h} (\autoref{fig:ds2_time_series}).
Although the density in the leg does not peak at the same time as the intensity, the fact that it peaks after the apex seems compatible with the simulations.

The velocities measured in datasets~1 and 8 globally match the predicted behavior.
Indeed, the downflows observed in dataset~1 all happen at the same time as the \linefexiimain intensity peaks, and all intensity peaks have associated downflows, except for the strong intensity peak at \SI{17}{\hour}.
In dataset~8, the two intensity peaks have associated downflows. However a third velocity peaks is seen at \SI{2.7}{\hour}, and does not appear to be associated with any intensity feature, which is puzzling.
Overall, the fact that most downflows happen at the same time as the corresponding intensity peaks is a strong clue that they are not instrumental artifacts.

Finally, we note that the AIA \SI{211}{\angstrom} intensity peaks before \SI{193}{\angstrom} in dataset~1, and that \SI{193}{\angstrom} peaks before \SI{171}{\angstrom} in dataset~2. This is consistent with previous reports that loops (undergoing TNE or not) are generally observed in their cooling phase \citep{WarrenEtAl2002, WinebargerEtAl2003, WinebargerWarren2005, Ugarte-UrraEtAl2006, Ugarte-UrraEtAl2009, Mulu-MooreEtAl2011, ViallKlimchuk2011, ViallKlimchuk2012, FromentEtAl2015}.

\section{Summary and conclusion} \label{sec:conclusion}

In order to detect velocity pulsations associated with long-period intensity pulsations, we used 11 sets of EIS rasters that correspond to the observation of known intensity pulsation events detected with AIA between 2010 and 2016 \citep{Froment2016}.
We detect velocity variations compatible with the expected pulsations in two of these datasets.
The variations are characterized by recurring downflows that happen at the same time as the intensity peaks.
The first dataset (1) contains six intensity peaks, four of which have matching velocity peaks.
The second dataset (8) contains two intensity peaks but shows three velocity peaks, with the third one occurring between the two intensity peaks.
Overall, we find a good, albeit not perfect, correlation between the observed intensity and velocity peaks for these two datasets.
The observed velocities are consistent with simulations from \citet{MikicEtAl2013} and \citet{FromentEtAl2017}, where strong downflows occur in one leg of the loop when the intensity peaks in the \SI{193}{\angstrom} channel of AIA.
Note that such velocity signature can correspond to condensation and evaporation cycles with or without formation of coronal rain.
The velocity variations have an amplitude of \num{4} and \SI{2.5}{\kmps} respectively, which is much lower than the $\sim\SI{30}{\kmps}$ flows produced in the simulations.
We argue that this difference is caused by the presence of emission from plasma at rest along the LOS, which decreases the amplitude of the measured velocity variation.
This also explains why we detect no velocity variations in the other datasets, which have a lower SNR combined with a lower intensity contrast, \ie more contamination from plasma outside of the pulsating loop.
Because the measured velocities are at the limits of the EIS capabilities, it is hard to know if the absence of detected velocity variations during some intensity peaks of dataset~1 indicate an absence of downflow in the loop, or simply lower velocities that fall below the detection threshold.

We also measured the density in the pulsating loops for two of the presented datasets.
Both show small density variations, which appear to be compatible with the behavior predicted by the simulations.
However, because these variations are faint ($\sim \SI{20}{\percent}$ in one dataset, and a single density peak the other), they do not provide a strong constraint to compare the simulations to the observations.

We detect velocity variations that are compatible with the pulsations predicted by the simulation. However, these pulsations are at the limits of the instrumental capabilities of EIS, and are therefore only detected in a fraction of observed events.
More observations are required in order to detect the pulsations without any ambiguity.
We have designed a new observation program for EIS, where we make the best compromise between cadence (one raster every 40 minutes), exposure time (\SI{30}{\second}), FOV ($\ang{;;304}\times\ang{;;512}$), and spatial resolution in the $X$ direction (\ang{;;4}).
The program has already been run once, but the observed active region contained no intensity pulsations. It is planed to run it again in the future.
This study highlights the need for a new generation of EUV spectrometers that can make observations with both high-SNR and high cadence at the same time.

\begin{acknowledgements}
The authors would like to thank the referee for contributing to the improvement of this paper.
We acknowledge P. Young for useful advice on the EIS data analysis.
\textit{Hinode} is a Japanese mission developed and launched by ISAS/JAXA, with NAOJ as domestic partner and NASA and STFC (UK) as international partners. It is operated by these agencies in co-operation with ESA and NSC (Norway).
AIA data are courtesy of NASA/SDO and the AIA science team.
This work used data provided by the MEDOC data and operations centre (CNES/CNRS/Univ. Paris-Sud), \href{http://medoc.ias.u-psud.fr}{medoc.ias.u-psud.fr}.
CHIANTI is a collaborative project involving George Mason University, the University of Michigan (USA), University of Cambridge (UK) and NASA Goddard Space Flight Center (USA).
We acknowledge support from the International Space Science Institute (ISSI), Bern, Switzerland to the International Team 401 “Observed Multi-Scale Variability of Coronal Loops as a Probe of Coronal Heating”.
C.F.: this research was supported by the Research Council of Norway, project no. 250810, and through its Centres of Excellence scheme, project no. 262622.
S.P. acknowledges the funding by CNES through the MEDOC data and operations centre.
\textit{Software:}
Astropy \citep{Astropy2013, Astropy2018},
ChiantiPy \citep{Dere2013-chiantipy},
SolarSoft \citep{FreelandHandy2012}.

\end{acknowledgements}

\begin{appendix}

\section{Monte-Carlo simulations of line fitting}\label{sec:mc_fit}

\begin{figure*}
\includegraphics[width=\textwidth]{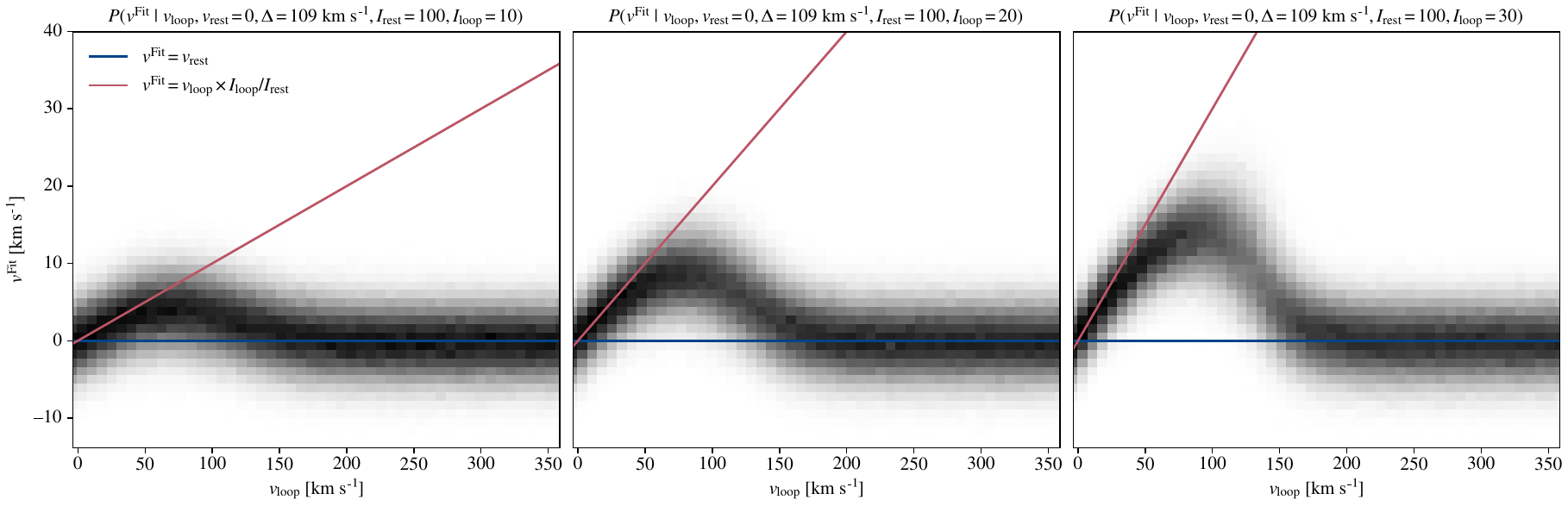}
\caption{
Stacked histograms showing dependency of $v^\text{Fit}$ on $v_\text{loop}$, for the fit of the single-gaussian function $G_1$ to synthetic 2-lines EIS spectra.
The three plots correspond to a SNR of 10, with (from left to right) intensity ratios $I_\text{loop} / I_\text{rest}$ of \SI{10}{\percent}, \SI{20}{\percent}, and \SI{30}{\percent}.
The blue lines represent the input position of the bright line, and the red lines show the empirical relation $v^\text{Fit} = v_\text{loop} \times I_\text{loop} / I_\text{rest}$.
Black bins corresponds to higher probabilities.
}
\label{fig:mc_ratio}
\end{figure*}

\begin{figure*}
\centering
\includegraphics[width=\textwidth]{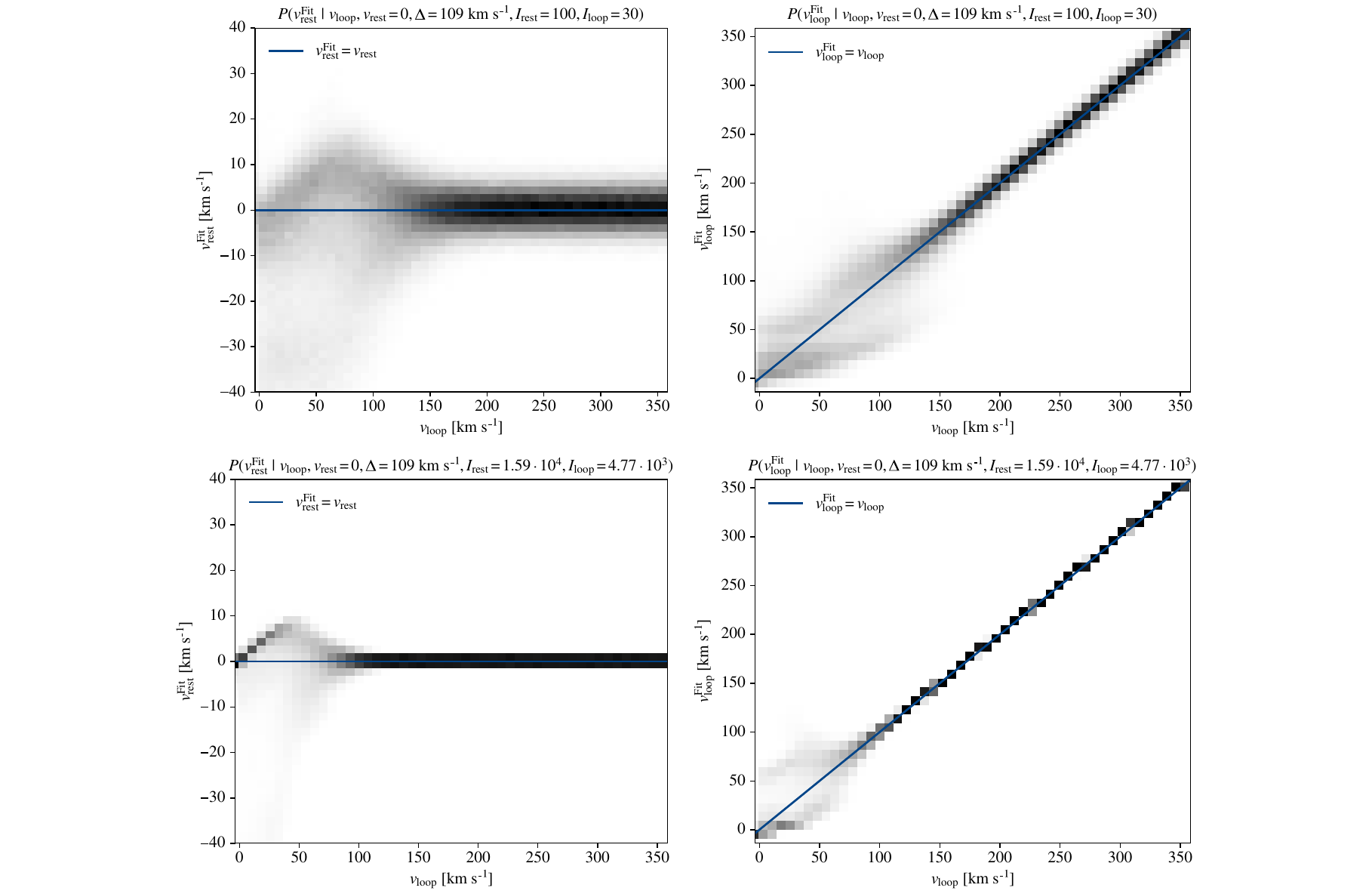}
\caption{
Stacked histograms for the fit of the double-gaussian function $G_2$ to synthetic 2-lines EIS spectra.
The left column shows the dependency of $v^\text{Fit}_\text{rest}$ on $v_\text{loop}$, and the right column the dependency of $v^\text{Fit}_\text{loop}$ on $v_\text{loop}$.
The top row corresponds to a SNR of \num{10}, while the bottom row corresponds to a SNR of \num{126}.
All spectra used for these plots had an intensity ratio $I_\text{loop} / I_\text{rest}$ of \SI{30}{\percent}.
The blue lines mark to the input position of the corresponding line (\ie $v_\text{rest}$ or $v_\text{loop}$).
Black bins corresponds to higher probabilities.
}
\label{fig:mc_dgb}
\end{figure*}

In order to better understand how the velocity from a faint line blended with an intense line can be retrieved, we perform Monte-Carlo simulations of line fitting.
To that end, we generate synthetic EIS spectra sampled every \SI{22}{\milli\angstrom} \citep{CulhaneEtAl2007}, and composed of two Gaussian line profiles with respective velocities $v_\text{loop}$ and $v_\text{rest}$, peak intensities $I_\text{loop}$ and $I_\text{rest}$, the same FWHM $\Delta$, and a global offset $b$.
(Note that in this appendix we assume that the line-of-sight is aligned with the loop, \ie $v_\text{loop} \cos\theta = v_\text{loop}$.)
The average number of photons as a function of the wavelength is therefore given by:

\begin{equation}
N_\lambda =
    I_\text{rest} \exp{ - 4\log{2} \left(\lambda - (v_\text{rest}/c + 1) \lambda_0\right)^2 / \Delta^2}
  + I_\text{loop} \exp{ - 4\log{2} \left(\lambda - (v_\text{loop}/c + 1) \lambda_0\right)^2 / \Delta^2}
  + b
  ,
\end{equation}

where $\lambda_0 = \SI{195.119}{\angstrom}$ is the rest wavelength of the simulated line, and $c$ is the speed of light.
We simulate photon noise by applying a realisation of the Poisson distribution, such that for a random variable $X$ and an integer $k$, $P(X = k | N_\lambda) = N_\lambda^k \exp{-N_\lambda} / k!$.

We then fit two model functions to these spectra: a single-gaussian function $G_1$, and a double-gaussian function $G_2$:

\begin{equation}
G_1(\lambda, I^\text{Fit}, v^\text{Fit}, \Delta^\text{Fit}, b^\text{Fit})
  = I^\text{Fit} \exp{ - 4\log{2} \left(\lambda - (v^\text{Fit}/c + 1) \lambda_0\right)^2 / {\Delta^\text{Fit}}^2} + b^\text{Fit}
\end{equation}

\begin{equation}
\begin{split}
G_2(\lambda, I_\text{rest}^\text{Fit}, I_\text{loop}^\text{Fit}, &v_\text{rest}^\text{Fit}, v_\text{loop}^\text{Fit}, \Delta^\text{Fit}_\text{rest}, \Delta^\text{Fit}_\text{loop}, b^\text{Fit}) = \\
  & I^\text{Fit}_\text{rest} \exp{ - 4\log{2} \left(\lambda - (v^\text{Fit}_\text{rest}/c + 1) \lambda_0\right)^2 / {\Delta^\text{Fit}_\text{rest}}^2} + \\
  & I^\text{Fit}_\text{loop} \exp{ - 4\log{2} \left(\lambda - (v^\text{Fit}_\text{loop}/c + 1) \lambda_0\right)^2 / {\Delta^\text{Fit}_\text{loop}}^2}
  + b^\text{Fit} \\
\end{split}
\end{equation}

For a given set of input parameters ($v_\text{rest}, v_\text{loop}, I_\text{rest}, I_\text{loop}, \Delta$), we generate \num{10000} spectra with different realizations of the noise, that we fit with both $G_1$ and $G_2$ in order to estimate the probability of retrieving each possible fit parameter values.
We explore different combinations of input parameters, in particular the position of the secondary line ($v_\text{loop}$), the ratio of the two lines ($I_\text{loop} / I_\text{rest}$), and the SNR (absolute value of $I_\text{rest}$).
We represent these results as stacked histograms (\autoref{fig:mc_ratio} and \autoref{fig:mc_dgb}), which show the probability to retrieve the values of a fit parameter, given the input parameters printed above the map, and the velocity of the secondary line $v_\text{loop}$ in abscissa.
Each column of these maps corresponds to the normalized histogram of the results of the \num{10000} fits performed for the corresponding input parameters.

In \autoref{fig:mc_ratio}, we present results of the fit of 2-lines spectra with the single-gaussian function $G_1$.
The three plots show the stacked histograms of $v^\text{Fit}$ as a function of $v_\text{loop}$, for different values of $I_\text{loop} / I_\text{rest}$ (\SI{10}{\percent}, \SI{20}{\percent}, and \SI{30}{\percent}), and a SNR of \num{10}.
For large separations between the two lines ($v_\text{loop} > \SI{150}{\kmps}$), the fitted Gaussian is centered on the brightest line of the spectrum, with $v^\text{Fit} = 0$.
However, for small separations ($v_\text{loop} < \Delta$), the centroid of the fitted Gaussian seems to follow the relation $v^\text{Fit} \lesssim v_\text{loop} \times I_\text{loop} / I_\text{rest}$.
Performing the same simulations with higher SNR values shows the same dependency of $v^\text{Fit}$ on $v_\text{loop}$, with lower dispersion.
Therefore, fitting a single Gaussian function to such 2-lines spectra yields information on the velocity of the weaker component, and the separation value can be computed with the knowledge of the intensity ratio between the two lines.

In \autoref{fig:mc_dgb}, we present stacked histograms that correspond to the fit of 2-lines spectra with the double-gaussian function $G_2$.
The left column shows the histograms of $v^\text{Fit}_\text{rest}$ as a function of $v_\text{loop}$, while the right column shows the histograms of $v^\text{Fit}_\text{loop}$ as a function of $v_\text{loop}$.
The top row corresponds to a SNR of \num{10}, and the bottom row to a SNR of \num{126}.
These SNR values are equivalent to respectively \num{0.4} and \SI{58}{\second} of integration time with the \ang{;;1} slit and in the \linefexiimain for typical active region count rates \citep[Table 12]{CulhaneEtAl2007}.
The maximum probability should be distributed around the blue line shown on each plot.
This is the case only for large separations of the two lines (\ie large values of $v_\text{loop}$).
For lower values ($v_\text{loop} < \SI{150}{\kmps}$ at a SNR of \num{10}, and $v_\text{loop} < \SI{100}{\kmps}$ at a SNR of \num{126}), the fit parameters are very disperse.
This shows that for line separations of less than one FWHM, it is not possible to accurately retrieve the velocity of the faint line, even with long exposure times.

\end{appendix}

\bibliographystyle{aa}
\bibliography{bibliography}

\end{document}